\documentclass[13pt, draftclsnofoot, onecolumn]{IEEEtran}

\pdfoutput=1

\usepackage{glossaries}

\usepackage{algorithm}
\usepackage{algpseudocode}
\usepackage{amsfonts}
\usepackage{tabularx}

\usepackage{xparse}
\newacronym{ecdf}{ECDF}{empirical cumulative distribution function}
\newacronym{tsnr}{TSNR}{transmit signal-to-noise ratio}
\newacronym{bcd}{BCD}{block coordinate descent}
\newacronym{bs}{BS}{base station}
\newacronym{ris}{RIS}{reconfigurable intelligent surface}
\newacronym{los}{LoS}{line-of-sight}
\newacronym{fcn}{FCN}{fully convolutional network}
\newacronym{snr}{SNR}{signal-to-noise ratio}
\newacronym{sinr}{SINR}{signal-to-interference-noise ratio}
\newacronym{ofdm}{OFDM}{orthogonal frequency-division multiplexing}
\newacronym{mmse}{MMSE}{minimum mean squared error}
\newacronym{mse}{MSE}{mean squared error}
\newacronym{mimo}{MIMO}{multiple-input-multiple-output}
\newacronym{wmmse}{WMMSE}{weighted minimum mean squared error}
\newacronym{cnn}{CNN}{convolutional neural network}
\newacronym{drl}{DRL}{deep reinforcement learning}
\newacronym{imm}{IMM}{interacting multiple model}
\newacronym{ppo}{PPO}{proximal policy optimization}
\newacronym{sac}{SAC}{soft actor-critic}
\newacronym{ekf}{EKF}{extended Kalman filter}
\newacronym{gae}{GAE}{generalized advantage estimation}
\newacronym{rl}{RL}{reinforcement learning}
\newacronym{kl}{KL}{Kullback–Leibler}
\newacronym{rsu}{RSU}{road-side unit}
\newacronym{crlb}{CRLB}{Cram\'er-Rao lower bound}
\newacronym{aoi}{AoI}{age of information}
\newacronym{voi}{VoI}{value of information}
\newacronym{kf}{KF}{Kalman filter}
\newacronym{pdf}{PDF}{probability density function}
\newacronym{ao}{AO}{alternating optimization}
\newacronym{wsr}{WSR}{weighted sum-rate}
\newacronym{mrt}{MRT}{maximum ratio transmission}
\newacronym{zf}{ZF}{zero-forcing}
\newacronym{ml}{ML}{machine learning}
\newacronym{iid}{i.i.d.}{independent and identically distributed}
\newacronym{csi}{CSI}{channel state information}

\newcommand{\mb}[1]{\mathbf{#1}}
\newcommand{\bs}[1]{\boldsymbol{#1}}
\newcommand{\tr}[1]{\text{tr}\left({#1}\right)}
\newcommand{\p}{\boldsymbol{\Psi}_{\theta}}
\newcommand{\pc}{\kappa}

\newcommand{\rev}[1]{\textcolor{red}{#1}}

\NewDocumentCommand{\DrawCubes}{O{} m m m m m m}{%
    \def\XGridMin{#2}
    \def\XGridMax{#3}
    \def\YGridMin{#4}
    \def\YGridMax{#5}
    \def\ZGridMin{#6}
    \def\ZGridMax{#7}
    \begin{scope}[canvas is xy plane at z=\ZGridMax]
      \draw [#1] (\XGridMin,\YGridMin) grid (\XGridMax,\YGridMax);
    \end{scope}
    \begin{scope}[canvas is yz plane at x=\XGridMax]
      \draw [#1] (\YGridMin,\ZGridMin) grid (\YGridMax,\ZGridMax);
    \end{scope}
    \begin{scope}[canvas is xz plane at y=\YGridMax]
      \draw [#1] (\XGridMin,\ZGridMin) grid (\XGridMax,\ZGridMax);
    \end{scope}
}%

\usepackage{tikz}
\usetikzlibrary{shapes.geometric, arrows}
\tikzstyle{box} = [rectangle, rounded corners, minimum width=2cm, minimum height=0.8cm,text centered, draw=black,]
\usepackage{subfigure}
\usepackage{hhline}
\usepackage{amsthm}
\usepackage{pgfplots}
\tikzstyle{box} = [rectangle, rounded corners, minimum width=2cm, minimum height=0.8cm,text centered, draw=black,]
\tikzstyle{label} = [minimum height=0.8cm,text centered, draw=none]

\usetikzlibrary{3d}
\makeatletter % from https://tex.stackexchange.com/a/48776/121799
\tikzoption{canvas is xy plane at z}[]{%
  \def\tikz@plane@origin{\pgfpointxyz{0}{0}{#1}}%
  \def\tikz@plane@x{\pgfpointxyz{1}{0}{#1}}%
  \def\tikz@plane@y{\pgfpointxyz{0}{1}{#1}}%
  \tikz@canvas@is@plane
}
\makeatother

\usepackage[backend=biber, style=ieee, url=false, isbn=false]{biblatex}
\bibliography{ref.bib}

\pgfplotsset{compat=1.17} 

\AtBeginBibliography{\small}

\begin{document}

\title{Reconfigurable Intelligent Surface Enabled Spatial Multiplexing with Fully Convolutional Network}

\author{Bile~Peng,~\IEEEmembership{Member,~IEEE},
        Ke~Guan,~\IEEEmembership{Senior Member,~IEEE},
        Danping~He,~\IEEEmembership{Member,~IEEE},
        Jan-Aike~Term\"ohlen,~\IEEEmembership{Student Member,~IEEE},
        Cong~Sun,~\IEEEmembership{Member,~IEEE},
        Tim Fingscheidt,~\IEEEmembership{Senior Member,~IEEE},
        and~Eduard~A.~Jorswieck,~\IEEEmembership{Fellow,~IEEE}% <-this % stops a space
\thanks{B. Peng, J. Term\"ohlen T. Fingscheidt, and E. A. Jorswieck are with Institute for Communications Technology, TU Braunschweig, 38106 Braunschweig, Germany (e-mail: \{peng,termoehlen,fingscheidt,jorswieck\}@ifn.ing.tu-bs.de).}% <-this % stops a space
\thanks{K. Guan and D. He are with State Key Laboratory of Rail Traffic Control and Safety, Beijing Jiaotong University, Beijing, China, 100044 (e-mail: ke.guan.cn@ieee.org, hedanping@bjtu.edu.cn).}
\thanks{C. Sun is with Beijing University of Posts and Telecommunications, Beijing, China, 100876 (e-mail: suncong86@bupt.edu.cn).}
}

\newtheorem*{theorem*}{Theorem}

\maketitle

\begin{abstract}
    \Gls{ris} is an emerging technology for future wireless communication systems.
    In this work, we consider downlink spatial multiplexing enabled by the \gls{ris} for \gls{wsr} maximization.
    In the literature, most solutions use alternating gradient-based optimization,
    which has moderate performance, high complexity, and limited scalability.
    We propose to apply a \gls{fcn} to solve this problem,
    which was originally designed for semantic segmentation \rev{and depth estimation} of images.
    The rectangular shape of the \gls{ris} and the spatial correlation of channels with adjacent \gls{ris} antennas due to the short distance between them encourage us to apply it for the \gls{ris} configuration.
    We design a set of channel features that includes both cascaded channels via the \gls{ris} and the direct channel.
    In the \gls{bs}, the differentiable \gls{mmse} precoder is used for pretraining
    and the \gls{wmmse} precoder is then applied for fine-tuning, which is nondifferentiable, more complex, but achieves a better performance.
    \rev{Furthermore, we propose a method to discretize the continuous phase shifts,
    which significantly reduces the requirement for hardware complexity with acceptable performance loss.}
    Evaluation results show that the proposed solution has higher performance, good robustness against channel estimation error and allows for a faster evaluation than the baselines.
    Hence it scales better to a large number of antennas,
    advancing the \gls{ris} one step closer to practical deployment.
\end{abstract}


\glsresetall

% \fi
\IEEEpeerreviewmaketitle

\section{Introduction}
\label{sec:intro}

\IEEEPARstart{T}{he} \gls{ris} is an emerging technology for next-generation wireless communication systems~\cite{bjornson2019massive,di2020smart,huang2020holographic}. 
It comprises many antennas on a surface.
Each antenna has the ability to receive a signal, process it without external power
(i.e., the signal cannot be amplified),
and reflect it.
With cooperation between the antennas,
the \gls{ris} can realize complicated signal processing.
Due to its simple structure and good compatibility with the other components of wireless communication systems (e.g., precoding in the \gls{bs}), 
the \gls{ris} is widely believed to be an essential part of the next-generation wireless communication systems and has been widely studied for optimizing 
\gls{wsr}~\cite{pan2020multicell,guo2020weighted},
capacity~\cite{zhang2021joint},
energy efficiency~\cite{huang2019reconfigurable}, reliability~\cite{besser2021reconfigurable},
physical layer security~\cite{chu2019intelligent}
and wireless power transfer~\cite{mishra2019channel,wu2019weighted}.

Among different applications of the \gls{ris}, we focus on improving the \gls{wsr} with spatial multiplexing in this work.
The high data rate is a major challenge for future wireless communication systems,
which can only be overcome by both acquiring broader bandwidth in higher frequencies
and improving the spectrum efficiency.
With a favorable propagation channel,
we can use the same resource block (a resource allocation unit in time and frequency domains) to serve multiple users and hence improve the spectrum efficiency.

Without \gls{ris}, the propagation channel depends mainly on positions of transmitter and receivers and nearby scatterers,
which we cannot optimize.
If we deploy an \gls{ris} in the environment,
we can optimize the environment by configuring the \gls{ris} such that the propagation channel is more favorable for spatial multiplexing.

This is a challenging problem because of its high dimensionality due to the large number of \gls{ris} antennas and the constraint that \gls{ris} does not amplify the signal.
In the literature, the \gls{mmse} precoder~\cite{Joham2005linear} and the \gls{wmmse} precoder~\cite{sampath2001generalized,christensen2008weighted,shi2011iteratively} are proposed for precoding in the \gls{bs} without consideration of the \gls{ris},
stochastic successive convex approximation~\cite{guo2020weighted},
majorization-maximization~\cite{huang2018achievable},
\gls{ao}~\cite{perovic2021maximum},
and
alternating direction method of multipliers (ADMM)~\cite{liu2021two}
are applied to jointly optimize the \gls{ris} configuration and the \gls{bs} precoding.
These proposed algorithms have achieved reasonable performance at the cost of high computational effort.
With similar problem formulations, 
spatial multiplexing in \gls{ris}-assisted uplink is considered and a gradient-based solution is proposed~\cite{Elmossallamy2021spatial}.
The Riemannian manifold conjugate gradient (RMCG) and the Lagrangian method are applied to configure multiple \glspl{ris} and \gls{bs} to serve users on the cell edge~\cite{li2020weighted}.
A joint precoding scheme with an \gls{ao} method is proposed for cell-free \gls{ris}-aided communication~\cite{zhang2021joint}.
The sum-rate is maximized with the majorization-maximization (MM) algorithm for grouped users in a \gls{ris}-aided communication system~\cite{zhou2020intelligent}.
% The zero-forcing precoding, majorization-maximization algorithm~\cite{huang2018achievable} and alternating maximization~\cite{huang2019reconfigurable} are applied to optimize the energy efficiency.
The downlink \gls{snr} is maximized with \gls{drl}~\cite{feng2020deep}.
The spectrum and energy efficiency are maximized with \gls{ris}~\cite{huang2019reconfigurable,gao2020reconfigurable}.
The interference caused by secondary networks is mitigated with an \gls{ris}~\cite{xu2020resource}.
The \gls{bs} configuration and the phase shifts of the \gls{ris} are iteratively optimized with the \gls{ofdm} transmission scheme~\cite{yang2020intelligent,li2020irs}.
The robust transmission scheme design is addressed as well~\cite{zhou2020framework}.

Although the above-mentioned pioneering works have achieved reasonably good performances compared to random phase shifts and scenarios without \gls{ris},
there is still considerable space for performance improvement 
and the computational complexities of these proposed algorithms are still too high for real-time application.
Besides, the high complexity constrains the number of \gls{ris} antennas.
In most works mentioned above, the number of \gls{ris} antennas is assumed to be less than 200,
which is far less than the vision of up to thousands of elements introduced in Ref.~\cite{di2020smart}.
In recent years, there have been attempts to apply machine learning to \gls{ris} optimization for better scalability.
\rev{For example, meta-learning is used to optimize the spectrum efficiency but only one user per \gls{ris} is considered in \cite{jung2021meta},
a reflection beamforming codebook is learnt with \gls{drl} for groups of users~\cite{zhang2021learning} (no spatial multiplexing is considered though),
the achievable rate is predicted with user location and beamforming in~\cite{sheen2021deep}.
}

In this work, we propose an approach based on \gls{fcn} and \gls{wmmse} precoding
that achieves better performance with less operating expenses. 
\rev{The \gls{fcn} is comprised of multiple convolutional layers and was first proposed in \cite{long2015fully} and
is successfully applied for semantic segmentation and depth estimation~\cite{arora2017fully},
where an image is fed into the neural network as input and the category (for semantic segmentation, categories are, e.g., the car, the road, or the sky) or the depth (for depth estimation, which is the distance between the camera and the object) of each pixel of the image is estimated as output.}
The first convolutional layer is able to extract the local features of the image 
while the following convolutional layers can process information on higher levels.
Inspired by the rectangular shape of the \gls{ris} 
and the spatial correlation of the channels of adjacent \gls{ris} antennas 
due to the close distance between the antennas~\cite{pizzo2020spatially},
we apply an \gls{fcn} for the \gls{ris} configuration.
In the literature, the \gls{fcn} has been applied to object detection~\cite{dai2016r}, 
visual tracking~\cite{wang2015visual} and
receiver design~\cite{honkala2021deeprx} besides the above-mentioned applications.
To the authors' best knowledge, it has not yet been applied to \gls{ris} configuration.

Our contributions in this paper are as follows:
\begin{itemize}
    \item We apply an \gls{fcn} for the \gls{ris} configuration. 
    We show that the \gls{fcn} is an efficient and scalable architecture for the considered \gls{wsr} maximization problem. 
    By introducing dropout layers, we prevent overfitting and improve the generalization of the trained models.
    \item We design a trick to consider the direct channel without \gls{ris} in the \gls{fcn},
    which makes the proposed approach more applicable because a weak direct channel exists in most use cases.
    \item We design a training process with the \gls{wmmse} precoder, 
    which is nondifferentiable and therefore incompatible with the gradient ascent optimization.
    By alternatingly training the \gls{fcn} and updating the precoding vectors, the \gls{fcn} can work properly with the \gls{wmmse} precoder.
    \rev{\item We propose a method to discretize the output phase shifts with low resolution (e.g., $\pi$ with binary, small codebook size) such that the solution can be applied to \gls{ris} with simple and cheap discrete phase shifts. The proposed method uses the Frobenius norm as a penalty and aims to minimize the performance loss compared to the continuous phase shift.}
\end{itemize}

In the following part of this work, Section~\ref{sec:problem} formulates the problem,
Section~\ref{sec:precoder} describes the precoding techniques,
Section~\ref{sec:solution} proposes our solution,
Section~\ref{sec:implementation} explains some issues of implementation and training and extends the method to discrete phase shifts,
Section~\ref{sec:results} presents the training and evaluation results and
Section~\ref{sec:conclusion} concludes the work.

\emph{Notations: } Throughout this work, we use the following notations:
$\mb{I}_n$ is an identity matrix of size $n \times n$, 
$\mb{0}_n$ is a zero matrix of size $n \times n$, 
$\mb{A} \star \mb{B}$ denotes the 2-dimensional cross-correlation between $\mb{A}$ and $\mb{B}$,
$|z|$ and $\arg(z)$ are the amplitude and phase of complex number $z$, respectively,
$\mb{A}^+$ denotes the pseudoinverse of matrix $\mb{A}$,
$\bar{\theta}$ refers to the untrainable neural network parameter set $\theta$ (i.e., it is considered constant during training),
$\exp(\mb{A})$ is the elementwise exponential function.
\section{Problem Formulation}
\label{sec:problem}

We consider an \gls{ris}-aided communication with direct propagation path between \gls{bs} and users. 
Our objective is to serve multiple users with the same resource block and to maximize the \gls{wsr} of the users.
The \gls{bs} is assumed to have multiple antennas and can do precoding subject to the transmit power constraint $E_{Tr}$.
The \gls{ris} is assumed to be rectangular with $H$ rows and $W$ columns of antennas.
Each \gls{ris} antenna has the ability to receive signal, adjust its complex phase without changing its amplitude and transmit it.

\begin{figure}[t]
    \centering
    \includegraphics[width=.45\textwidth,bb=0 0 624 384]{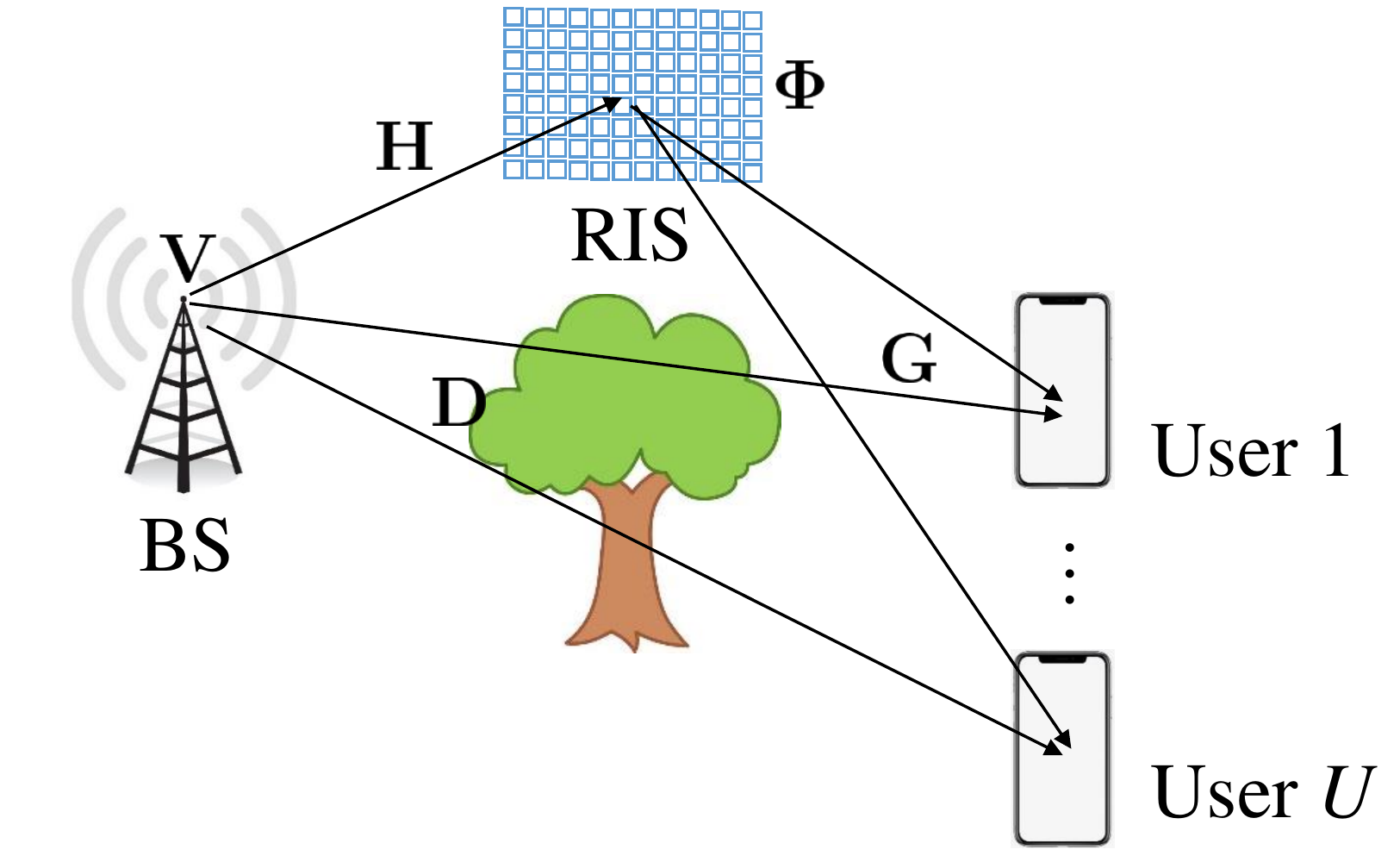}
    \caption{System model.}
    \label{fig:system_model}
\end{figure}

We denote the precoding matrix as $\mb{V}$ of size $M\times U$, where $M$ is the number of \gls{bs} antennas,
$U$ is the number of single-antenna users,
the channel from \gls{bs} to \gls{ris} as $\mb{H}$ of size $N\times M$, 
where $N$ is the number of \gls{ris} antennas ($N=WH$) and element $h_{nm}$ in row $n$ and column $m$ represents the channel gain from the $m$th antenna of the \gls{bs} to the $n$th antenna of the \gls{ris}.
The \gls{ris} signal processing is denoted by the diagonal matrix $\boldsymbol{\Phi}$ of size $N\times N$, 
where the element $\phi_{nn}$ in row $n$ and column $n$ is $e^{j\psi_n}$, 
with $\psi_n$ being the phase shift of antenna $n$. 
We consider both continuous and discrete phase shift in this work.
If the phase shift is continuous, $\psi_n \in [0, 2\pi)$.
If the phase shift is discrete, $\psi_n \in \boldsymbol{\Phi}_d$,
where $\boldsymbol{\Phi}_d$ is the set of predefined discrete phase shifts.
The channel matrix from \gls{ris} to users is denoted as $\mb{G}$ of size $U\times N$, 
where the element $g_{un}$ denotes the channel gain from \gls{ris} antenna $n$ to user $u$. 
The direct channel from \gls{bs} to users is denoted as $\mb{D}$ of size $U\times M$, 
where the element $d_{um}$ in row $u$ and column $m$ is the channel gain from the $m$th \gls{bs} antenna to the $u$th user, the transmission is described as
\begin{equation}
\mb{y} = \left(\mb{G} \boldsymbol{\Phi} \mb{H} + \mb{D} \right) \mb{V} \mb{x} + \mb{n},
\label{eq:transmission_los}
\end{equation}
where $\mb{x}$ is the vector of transmitted symbols, 
$\mb{y}$ is the vector of received symbols and $\mb{n}$ is the vector of thermal noise. 
All three vectors have the same size of $U \times 1$. 
The whole system model is illustrated in Fig.~\ref{fig:system_model}.

Let 
\begin{equation}
\mb{C} = (\mb{G} \boldsymbol{\Phi} \mb{H} + \mb{D}) \mb{V}
\label{eq:channel}
\end{equation}
and $c_{uv}$ being the element of $\mb{C}$ in row $u$ and column $v$, the objective is to maximize the \gls{wsr}. 
Therefore, the problem can be formulated as
\begin{equation}
\begin{array}[t]{ll}
    \max_{\mb{V}, \boldsymbol{\Phi}} & %\begin{aligned}
 f = \sum_{u=1}^U\alpha_u\log_2\left(1+\frac{c_{uu}^2}{\sum_{v\neq u}c_{uv}^2+\frac{1}{\rho}}\right)
%\end{aligned} 
\\
    \text{s.t.} & \tr{\mb{V}\mb{V}^H} \leq E_{Tr}\\
    & |\phi_{nn}|=1\\
    & |\phi_{nn'}|=0 \text{ for } n \neq n',
\end{array}
\label{eq:problem}
\end{equation}
where $\alpha_u$ is weight of user~$u$, $\alpha_u \in [0, 1]$ and $\sum_{u=1}^U\alpha_u = 1$, and $\rho$ is the \gls{tsnr},
which is the ratio between transmit power and noise power.
\section{The MMSE and WMMSE precoder}
\label{sec:precoder}

In problem \eqref{eq:problem}, $\mb{H}$, $\mb{G}$ and $\mb{D}$ are given and we would like to optimize both $\mb{V}$ and $\boldsymbol{\Phi}$. 
While the \gls{ris} optimization problem (i.e., optimizing $\boldsymbol{\bs{\Phi}}$) is new, the precoding problem (i.e., optimizing $\mb{V}$) has been intensively studied in the literature.
Different precoding techniques, such as \gls{mrt}, \gls{zf}, \gls{mmse} and \gls{wmmse} are proposed.
Among them, the \gls{mmse} precoder can minimize the mean squared error caused by both interference and noise with a closed form solution and is the optimal precoder for a single user~\cite{Joham2005linear}.
The iterative \gls{wmmse} precoder introduces weights of the \glspl{mse} of different users in the multi-user \gls{mimo} system.
It is proven that minimizing the sum of weighted \glspl{mse} for certain weights is equivalent to maximizing the \gls{wsr}~\cite{shi2011iteratively}.
However, the \gls{wmmse} precoder does not have a closed-form solution and is only available as an iterative algorithm, which significantly increases its complexity compared to the \gls{mmse} precoder. 
Accordingly we cannot compute its derivative. 
In the following, we briefly introduce the two precoders that will be applied jointly with the \gls{ris} optimization.

\subsection{The MMSE Precoder}
\label{sec:mmse}

In the \gls{mmse} precoder, the error is defined as the difference between transmitted symbols $\mb{x}$ and received symbols $\mb{y}$ scaled by a proper factor $\beta^{-1}$ (because the received symbols are weaker due to the propagation loss).
Accordingly, the \gls{mse} is defined as
\begin{equation}
    \begin{aligned}
    \mb{e}= & \mathbb{E}( || \beta^{-1}\mb{y}-\mb{x} ||^2_2 )\\
    = & \mathbb{E}\left(||\beta^{-1} (\mb{C}\mb{V}\mb{x} + \mb{n}) - \mb{x}||_2^2\right).
    \end{aligned}
    \label{eq:error_mmse}
\end{equation}
The objective is to minimize the \gls{mse} subject to the transmit power constraint:
\begin{equation}
\begin{array}{rl}
    \min_{\beta, \mb{V}} & \mathbb{E}\left(||\beta^{-1} (\mb{C}\mb{V}\mb{x} + \mb{n}) - \mb{x}||_2^2\right) \\
    \text{s.t.} & \tr{\mb{V}\mb{V}^H} \leq E_{Tr}. \\
\end{array}
\label{eq:mmse_formulation}
\end{equation}

When we assume $\mathbb{E}(\mb{x}\cdot \mb{x}^T)=\mb{I}_2$, $\mathbb{E}(\mb{n}\cdot \mb{n}^T)=1/\rho \cdot \mb{I}_2$ and $\mathbb{E}(\mb{n}\cdot \mb{x}^T)=\mb{0}_2$, we obtain a closed-form solution~\cite{Joham2005linear}
\begin{equation}
\mb{V} = \beta\left(\mb{C}^H\mb{C}+\frac{1}{\rho}\mb{I}\right)^{-1}\mb{C}^H
\label{eq:mmse}
\end{equation}
where
\begin{equation}
\beta=\sqrt{\frac{E_{Tr}}{\tr{\left(\mb{C}^H\mb{C}+\frac{1}{\rho}\mb{I}\right)^{-2}\mb{C}^H\mb{C}}}}.
\end{equation}

The \gls{mmse} precoder is the optimal precoder to maximize the data rate for a single user.
However, it uses the same scalar $\beta^{-1}$ to scale all received symbols of multiple users.
If the received symbols at different users have very different signal strengths due to different channel gains,
the \gls{mmse} precoder is forced to equalize the channel gains,
which results in a suboptimal performance.
Besides, it minimizes the sum of mean squared errors of all users.
If we would like to optimize the \gls{wsr},
the weights (i.e., $\alpha_u$ in \eqref{eq:problem}) cannot be considered in the \gls{mmse} precoder.
Therefore, the \gls{mmse} precoder is not the optimal precoder to maximize the \gls{wsr} with multiple users.
Nevertheless, it is still a good choice due to its satisfying performance and simple, closed form (and therefore differentiable) solution.

\subsection{The WMMSE Precoder}
\label{sec:wmmse}

Similar to the \gls{mmse} precoder, we define the error of user~$u$ as
\begin{equation}
    e_u = \xi_u y_u - x_u,
    \label{eq:wmmse_error}
\end{equation}
where $\xi_u$ is the scaling factor of user~$u$ (corresponding to $\beta^{-1}$ in \eqref{eq:error_mmse} but can be different for different users),
$y_u$ is the received symbol of user~$u$ (i.e., the $u$th element of $\mb{y}$ in \eqref{eq:transmission_los})
and $x_u$ is the transmitted symbol of user~$u$ (i.e., the $u$th element of $\mb{x}$ in \eqref{eq:transmission_los}).

The \gls{wmmse} precoder considers the following problem
\begin{equation}
\begin{array}{rl}
\min_{w_u, \mb{v}_u, \xi_u} & \sum_{u=1}^U\ \alpha_u (w_u \mathbb{E}(e_u e_u^H) - \log w_u)\\
\text{s.t.} & \sum_{u=1}^U \tr{\mb{v}_u \mb{v}_u^H} \leq E_{Tr},
\end{array}
\label{eq:wmmse_formulation}
\end{equation}
where $w_u$ is the weight of the \gls{mse} of user~$u$,
$U$ is the number of users,
$\mb{v}_u$ is the precoding vector for user~$u$ of size $M \times 1$ (i.e., the $u$th column of $\mb{V}$ in \eqref{eq:transmission_los}).
It is proven that problem~\eqref{eq:wmmse_formulation} is equivalent to the problem of \gls{wsr} maximization \eqref{eq:problem} (when $\bs{\Phi}$ is fixed) in the sense that their optimal solutions of $w_u$ and $\mb{v}_u$ are the same when $\xi_u$ is carefully selected for all users $u$. 
% Therefore, the \gls{wmmse} precoder is the optimal precoder to maximize the \gls{wsr}.

Unfortunately, there is no closed-form solution to \eqref{eq:wmmse_formulation}. 
Instead, Algorithm~\ref{alg:wmmse} is proposed to iteratively solve problem~\eqref{eq:wmmse_formulation},
where we define $\mb{c}_u$ as the channel to user~$u$ of size $1 \times M$ (i.e., the $u$th row of $\mb{C}$),
$\mu^*$ is the Lagrange multiplier of the power constraint and 
is chosen such that the transmit power constraint in \eqref{eq:wmmse_formulation} is satisfied.

\begin{algorithm}
\caption{Iterative WMMSE Precoder}
\label{alg:wmmse}
\begin{algorithmic}
\State Initialize $\mb{v}_u$
\Repeat
\State $w_u' \leftarrow w_u$ for all users $u$
\State $\xi_u \leftarrow \left( \sum_{\nu=1}^U \mb{c}_u \mb{v}_\nu \mb{v}_\nu^H \mb{c}_u^H + 1/\rho \right)^{-1} \mb{c}_u \mb{v}_u$ for all users $u$
\State $w_u \leftarrow \left( 1 - \xi_u^H \mb{c}_u \mb{v}_u \right)^{-1}$ for all users $u$
\State $\mb{v}_u \leftarrow \alpha_u \left( \sum_{\nu=1}^U \alpha_\nu \mb{c}_\nu^H\xi_\nu w_\nu \xi_\nu^H \mb{c}_\nu + \mu^* \mb{I}_M \right)^{-1} \mb{c}_u^H \xi_u w_u$ for all users $u$
\Until{$\big| \sum_{u=1}^U |w_u| - \sum_{u=1}^U |w'_u| \big|\leq \epsilon$}
\end{algorithmic}
\end{algorithm}

The \gls{wmmse} precoder is the optimal precoder to maximize the \gls{wsr} \rev{(see Theorem~1 in \cite{shi2011iteratively} for details)}. However, its complexity is considerably higher than the \gls{mmse} precoder. Besides, the iterative solution is not differentiable.
\section{Proposed FCN Based Approach}
\label{sec:solution}

\subsection{The FCN Approach}

% An \gls{fcn} contains multiple two-dimensional convolutional layers, 
% where one convolutional layer processes information on a plane. 
% For example, in semantic segmentation,
% the input is an image and the elementary unit of the image is a pixel.
% Each pixel has a color characterized by three numbers (e.g., red, green and blue),
% resulting in three so-called \emph{features}, one for each color. 
An \gls{fcn} consists of several two-dimensional convolutional layers,
where each layer processes the feature maps of its preceding layer and extracts new feature maps from them.
For example, in semantic segmentation \rev{and depth estimation}, usually an RGB image is used as input for the network. 
The filters in the first layer operate on the three color channels of the image and extract feature maps for the following convolutional layer.
The filter kernels learned in the first layer are three-dimensional as they extract features from a three-dimensional input tensor.
Each filter kernel generates one output feature map.
The output of a semantic segmentation \gls{fcn} in a certain task is a category for each pixel (e.g., road, car, pedestrian)
\rev{wheres the output of a depth segmentation \gls{fcn} is the estimated depth for each pixel}.
The channels $\mb{H}$ and $\mb{G}$ reveal spatial correlation due to the short distance between \gls{ris} antennas.
This property shows similarity to images, where adjacent pixels form an object,
which can be captured by a two-dimensional convolutional filter,
\rev{thereby justifying the choice of an \gls{fcn} architecture}.
\rev{The optimal output of each \gls{ris} antenna depends on both its own feature and features of other antennas, 
which is the same to semantic segmentation and depth estimation\footnote{\rev{Note that the neural network is an optimized mapping from channel features to the \gls{ris} phase shifts. 
For different channels as inputs, 
the phase shifts are different.
Therefore, the achieved data rate is not the ergodic rate.}}.}
Encouraged by these analogies, we would like to compute $\bs{\Phi}$ with the \gls{fcn}. 
The elementary unit of an \gls{ris} is an \gls{ris} antenna. 
Its features are the properties characterizing the wireless channels with respect to this \gls{ris} antenna, which will be defined later in Section~\ref{sec:definition_features}.
Other conventional neural network architectures, such as neural networks with dense layers would not work for such high-dimensional input and output.

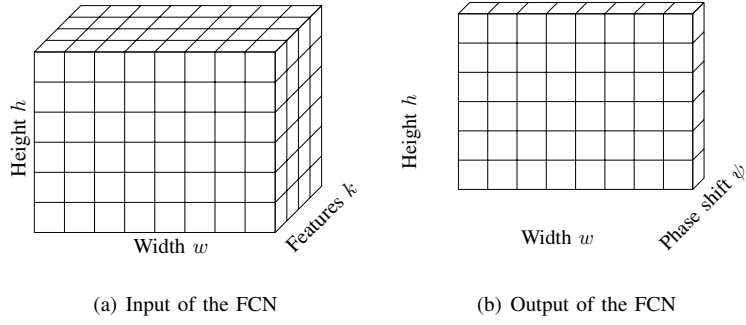
\begin{figure}[t]
    \centering
    \subfigure[Input of the FCN]{
    \resizebox{.3\textwidth}{!}{\begin{tikzpicture}
   y={(0.5cm,0.25cm)},x={(0.5cm,-0.25cm)},z={(0cm,{veclen(0.5,0.25)*1cm})}
    ]
    \DrawCubes [step=5mm,thin]{0}{4}{0}{3}{0}{2}
    \node (width) [rectangle, yshift=-1cm, xshift=1.5cm] {Width $w$};
    \node (height) [rectangle, yshift=.8cm, xshift=-1cm,rotate=90] {Height $h$};
    \node (channels) [rectangle, yshift=-.5cm, xshift=4cm,rotate=45] {Features $k$};
\end{tikzpicture}}}
    \subfigure[Output of the FCN]{
    \resizebox{.3\textwidth}{!}{\begin{tikzpicture}
   y={(0.5cm,0.25cm)},x={(0.5cm,-0.25cm)},z={(0cm,{veclen(0.5,0.25)*1cm})}
    ]
    \DrawCubes [step=5mm,thin]{0}{4}{0}{3}{0}{0.5}
    \node (width) [rectangle, yshift=-1cm, xshift=1.5cm] {Width $w$};
    \node (height) [rectangle, yshift=.8cm, xshift=-1cm,rotate=90] {Height $h$};
    \node (channels) [rectangle, yshift=-.5cm, xshift=4cm,rotate=45] {Phase shift $\psi$};
\end{tikzpicture}
  }}
    \caption{Input and output format of the FCN. The input is a three-dimensional array with dimensions of channel features, height and width. Channel feature $k$ of RIS antenna at position $(h, w)$ is stored in coordinate $(k, h, w)$. The output is a two-dimensional array with dimensions of height and width. The phase shift of RIS antenna at position $(h, w)$ is stored in coordinate $(h, w)$.}
    \label{fig:format_input_output}
\end{figure}

As shown in Fig.~\ref{fig:format_input_output}(a), the input of the \gls{fcn} is a three-dimensional array,
where the first dimension is the feature map index $k \in \{1, \dots, K \}$ with $K$ the number of features, 
second and third dimensions are height $h \in \{ 1, \dots, H\}$ and width $w \in \{ 1, \dots, W\}$, respectively. 
For example, the value of feature map index~$k$ of the \gls{ris} antenna at position $(h, w)$ is stored in coordinate $(k, h, w)$ in the three-dimensional array.
The output of the \gls{fcn} is a two-dimensional array of the same height and width as the input, 
but with only one feature map being the phase shift of the antenna,
as shown in Fig.~\ref{fig:format_input_output}(b).

Define $\mb{Q}_{\tau}$ as the $\tau$th feature map of the output of a convolutional layer. 
It is computed as
\begin{equation}
    \mb{Q}_{\tau} = b_{\tau} + \sum_{k=1}^K \mb{W}_{k\tau} \star \mb{P}_k,
    \label{eq:conv2d}
\end{equation}
where $b_{\tau}$ is the bias of feature map with index $\tau$, 
$\mb{W}_{k\tau}$ is the filter kernel which takes feature map with index~$k$ as input and produces feature map with index~$\tau$ as output,
$\mb{P}_k$ is the array of feature map with index~$k$ of the input,
$K$ is the number of feature maps of the input.

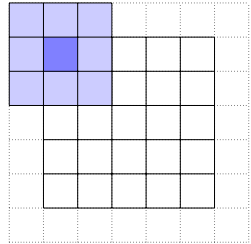
\begin{figure}[htbp]
    \centering
    \resizebox{.2\textwidth}{!}{
    \begin{tikzpicture}
    \draw[step=1cm] (0,0) grid (5, 5);
    \draw[step=1cm, dotted] (-1,-1) grid (6, 6);
    \fill[blue!20!white] (-1,3) rectangle (2,6);
    \fill[blue!50!white] (0,4) rectangle (1,5);
    \draw[step=1cm] (-1,3) grid (2, 6);
    \end{tikzpicture}
    }
    \caption{Illustration of filter and zero padding. 
    A filter of size $3\times 3$ (the light blue area) process information in the units it covers according to \eqref{eq:conv2d} and stores the result in the dark blue unit in the output of the layer.
    Since the filter size is $3\times 3$, the output size is reduced by 2 in both width and height.
    In order to keep the output size identical to the input size,
    we add zeros around the actual input (dotted units around the solid units),
    which is referred to \emph{zero padding}.
    The zeros are also in accordance with reality: since the channel features are amplitudes and phases.
    For units without an antenna (the dotted units),
    the amplitude is 0 and the phase does not matter.
    }
    \label{fig:padding}
\end{figure}

When applied in a straightforward manner, the cross-correlation operation makes the size of $\mb{Q}_{\tau}$ smaller than the size of $\mb{P}_k$. 
We use zero padding~\cite{albawi2017understanding} such that the size of output equals the input size,
which is illustrated in Fig.~\ref{fig:padding}.
Intuitively, the number of zeros added to each side of the input in the horizontal direction is
\begin{equation}
    w_z = \frac{w_f - 1}{2}
\end{equation}
where $w_f$ is the filter size in the horizontal direction.
Note that $w_f$ has to be an odd number such that $w_z$ is an integer.
A similar relationship can be obtained for the vertical direction.
In this way, the sizes of input and output are kept identical.

\begin{figure}[htbp]
    \centering
    \resizebox{.2\textwidth}{!}{
    \begin{tikzpicture}
\fill[blue!10!white] (1, 1) rectangle (8, 8);
\fill[blue!30!white] (2, 2) rectangle (7, 7);
\fill[blue!60!white] (3, 3) rectangle (6, 6);
\fill[blue] (4,4) rectangle (5,5);
\draw[step=1cm] (0,0) grid (9, 9);
\end{tikzpicture}}
    \caption{Illustration of receptive field in different FCN layers. We consider the darkest unit in the middle and a filter size of $3\times 3$. After the first convolutional layer, the unit under consideration is in the receptive field of the units in the second darkest areas. After the second convolutional layer, the unit under consideration is in the receptive field of the units in the lighted areas. In this way, the information in the considered unit propagates to all units through 4 layers.}
    \label{fig:receptive_field}
\end{figure}
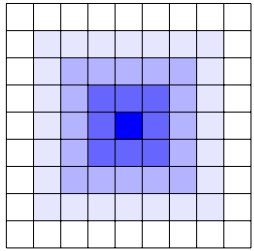

\rev{The size of the filters and the number of convolutional layers should be designed in such a way, 
that each output elementary unit has its receptive field\footnote{The region of the input that produces the output} of the whole \gls{ris}.
This idea is illustrated in Fig.~\ref{fig:receptive_field}.
% For example, if the \gls{ris} has the size of 16 $\times$ 16, 
% a filter of size 5 $\times$ 5 would have a receptive field with largest distance of $(5-1)/2=2$ in each direction (i.e., left, right, up and down) in the input of the same layer.
% Therefore, we need 8 convolutional layers such that the output elementary unit in one corner has access to the input elementary unit in the opposite corner for an \gls{ris} size of $16\times 16$.
Since every filter aggregates information from both sides in the horizontal direction by $(w_f - 1) / 2$,
the required number of layers is
\begin{equation}
    N_{\text{layers}} = \frac{(w_f - 1) / 2}{W},
    \label{eq:number_layers}
\end{equation}
such that the receptive field of a unit on one edge covers a unit on the opposite edge.
Equation \eqref{eq:number_layers} can be extended to the vertical layer easily.}

In order to prevent overfitting and to improve robustness, we put a dropout layer~\cite{hinton2012improving} between each convolutional layer,
which randomly turns off a certain portion of neurons.

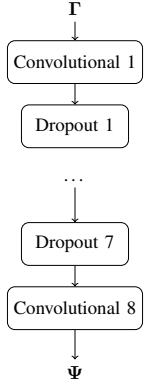
\begin{figure}[t]
    \centering
    \resizebox{.12\textwidth}{!}{
    \begin{tikzpicture}
\node (gamma) [rectangle] {$\bs{\Gamma}$};
\node[box, below of=gamma] (conv1) {Convolutional 1};
\node[box, below of=conv1, yshift=-.2cm] (dropout1) {Dropout 1};
\draw[->] (gamma) -- (conv1);
\draw[->] (conv1) -- (dropout1);
\node[rectangle, below of=dropout1, yshift=-.0cm] (dots) {\dots};
 \node[box, below of=dots, yshift=-.2cm] (dropout7) {Dropout 7};
\node[box, below of=dropout7, yshift=-.2cm] (conv8) {Convolutional 8};
\draw[->] (dots) -- (dropout7);
\draw[->] (dropout7) -- (conv8);
\node[rectangle, below of=conv8, yshift=-.2cm] (psi) {$\mb{\Psi}$};
\draw[->] (conv8) -- (psi);
\end{tikzpicture}
    }
    \caption{Structure of the proposed FCN for phase shift optimization.}
    \label{fig:fcn_structure}
\end{figure}

After all layers, the output of the \gls{fcn} $\bs{\Psi}$ is the array of phase shifts of all \gls{ris} antennas of size $H\times W$.
The structure of the proposed \gls{fcn} is shown in Fig.~\ref{fig:fcn_structure}.
The computation of the \gls{fcn} is denoted as $\bs{\Psi} = \p(\bs{\Gamma})$, 
where $\theta$ is the parameters of the \gls{fcn} (i.e., all biases and filters of all layers),
$\bs{\Gamma}$ is the input feature map,
$\bs{\Psi}=(\psi_n)$ is the two-dimensional array with linearly addressed elements $\psi_{n}, n=1, \dots, N$, $n$ is the index of the \gls{ris} antenna. 
For antenna at position $(h, w)$, we have $n = H \cdot (h - 1) + w$.

\subsection{Definition of Features}
\label{sec:definition_features}
\rev{In problem~\eqref{eq:problem},
we assume that the \gls{csi} at the \gls{ris} controller is known, which is a common assumption in relevant works~\cite{yu2020robust,fu2019intelligent,hou2020reconfigurable,zhu2020power,guo2020weighted}.
Although the channel estimation involving an \gls{ris} is challenging due to the large number of \gls{ris} antennas,
there are multiple promising attempts to estimate the channel with sufficient accuracy in real-time, such as \cite{nadeem2019intelligent,he2019cascaded,taha2021enabling,he2021channel}.
Besides, the \gls{csi} can also be used as intermediate information to derive an \gls{ris} optimization method with an input easier to obtain than the \gls{csi} itself, 
such as user locations~\cite{fascista2022ris}.
Therefore, we believe that \gls{csi} based \gls{ris} optimization has practical importance in the near future.}

Among the channel matrices,
$\mb{H}$ is assumed to be constant because both \gls{bs} and \gls{ris} are stationary,
$\mb{G}$ and $\mb{D}$ vary with the user positions and are required to compute $\bs{\Phi}$ and $\mb{V}$. 
We would like to interpret $\mb{G}$ and $\mb{D}$ as two feature maps with element index $n$ referring to a certain \gls{ris} antenna. Accordingly, we can use our \gls{fcn}. 
While each column of $\mb{G}$ can be injectively mapped to each \gls{ris} antenna, 
$\mb{D}$ is irrelevant to the \gls{ris}. Therefore, we define $\mb{J}=\mb{D}\mb{H}^+$ and \eqref{eq:transmission_los} becomes
\begin{equation}
\mb{y} = \left(\mb{G} \boldsymbol{\Phi} + \mb{J}\right) \mb{H} \mb{V} \mb{x} + \mb{n}.
\label{eq:transmission_los_equivalent}
\end{equation}
Equation \eqref{eq:transmission_los_equivalent} can be interpreted as follows:
the precoded signal $\mb{Vx}$ is transmitted through channel $\mb{H}$ to every \gls{ris} antenna and then through channel $\mb{G}\bs{\Phi} + \mb{J}$ to the users. 
Both $\mb{G}$ and $\mb{J}$ have $N$ columns and their columns can be injectively mapped to the \gls{ris} antennas.
Therefore, the features of an \gls{ris} element at position $(w, h)$ can be defined as
\begin{equation}
\begin{aligned}
    \bs{\gamma}_{wh} = ( & |g_{1n}|, \arg(g_{1n}), \dots, |g_{Un}|, \arg(g_{Un}),\\
    & |j_{1n}|, \arg(j_{1n}),\dots, |j_{Un}|, \arg(j_{Un})),
\end{aligned}
\label{eq:feature_definition}
\end{equation}
where $g_{k\tau}$ is the element in row~$k$ and column~$\tau$ of $\mb{G}$,
$j_{k\tau}$ is defined similarly to matrix $\mb{J}$.
In this way, we use 4$U$ features to characterize the wireless channels with $U$ users with respect to each \gls{ris} antenna~$n$, yielding the \gls{fcn} input $\bs{\Gamma}=(\mb{G}, \mb{J})$.

\subsection{Objective Functions}

Our objective function~$f$ in \eqref{eq:problem} is a function of channels $\mb{G}$ and $\mb{D}$, 
\gls{ris} phase shifts $\bs{\Phi}$ and \gls{bs} precoding $\mb{V}$. 
When $\bs{\Phi}$ is determined, 
$\mb{V}$ can be computed with either \gls{mmse} precoder or \gls{wmmse} precoder as described in Section~\ref{sec:precoder},
which are denoted as $\mb{V}_{\text{MMSE}}(\mb{G}, \mb{D}, \bs{\Phi})$ and $\mb{V}_{\text{WMMSE}}(\mb{G}, \mb{D}, \bs{\Phi})$, respectively.
It is to note that $\mb{V}_{\text{MMSE}}$ is a differentiable function but $\mb{V}_{\text{WMMSE}}$ is not differentiable.

Since $f$ is a function of $\mb{G}$, $\mb{D}$, $\bs{\Phi}$ and $\mb{V}$,
we can express $f$ as $f(\mb{G}, \mb{D}, \bs{\Phi}, \mb{V})$. Therefore, for \gls{mmse} precoding, the objective function is
\begin{equation}
\begin{aligned}
    f\bigg(& \mb{G}, \mb{D}, \exp\big(j\p(\mb{G}, \mb{D})\big), \\
    & \mb{V}_{\text{MMSE}}\Big(\mb{G}, \mb{D}, \exp\big(j\p(\mb{G}, \mb{D})\big)\Big)\bigg).
\end{aligned}
\label{eq:objective_mmse}
\end{equation}

For \gls{wmmse} precoding, the objective function is
\begin{equation}
\begin{aligned}
    f\bigg(& \mb{G}, \mb{D}, \exp\big(j\p(\mb{G}, \mb{D})\big), \\
    & \mb{V}_{\text{WMMSE}}\Big(\mb{G}, \mb{D}, \exp\big(j\bs{\Psi}_{\bar{\theta}}(\mb{G}, \mb{D})\big)\Big)\bigg).
\end{aligned}
\label{eq:objective_wmmse}
\end{equation}
Please note that $\bs{\Psi}_{\bar{\theta}}$ in \eqref{eq:objective_wmmse} is untrainable, i.e., $\bar{\theta}$ is given and $\bs{\Psi}_{\bar{\theta}}(\mb{G}, \mb{D})$ is a constant.
Therefore, we do not need to compute the derivative of $\mb{V}_{\text{WMMSE}}$ since it does not contain trainable variable $\theta$.

\begin{figure}[htbp]
    \centering
    \begin{footnotesize}
    \subfigure[With MMSE precoder]{\resizebox{.45\textwidth}{!}{\begin{tikzpicture}[cross/.style={path picture={ 
  \draw[black]
(path picture bounding box.south east) -- (path picture bounding box.north west) (path picture bounding box.south west) -- (path picture bounding box.north east);
}}]
\node (G) [rectangle] {$\mb{G}$};
\node (fcn) [box, below of=G, xshift=.3cm, yshift=-1.5cm] {FCN};
\draw [->] (G) -- ([xshift=-.3cm]fcn.north);
\node (phi) [rectangle, below of=fcn] {$\bs{\Phi}$};
\draw [->] (fcn) -- (phi);

\node (precoder) [box, right of=fcn, xshift=2cm] {MMSE Precoder};
\draw [->] (G) -| ([xshift=-.3cm]precoder.north);
\draw [->] (phi) -| (precoder);
% \node (W) [rectangle, right of=precoder, xshift=1cm] {$\mb{V}$};
% \draw [->] (precoder) -- (W);

\node (sum_rate) [box, right of=precoder, xshift=2cm] {WSR};
\draw [->] (G) -| ([xshift=-.2cm]sum_rate.north);
\draw [->] (phi) -| (sum_rate);
% \draw [->] (precoder) -- (sum_rate);
\draw [->] (precoder) -- node [midway,above ] {$\mb{V}$} (sum_rate);

\node (D) [rectangle, above of=fcn, yshift=1.0cm, xshift=0.3cm] {$\mb{D}$};
\node (H+) [rectangle, above of=fcn, yshift=0.cm, xshift=1.2cm] {$\mb{H}^+$};
\node (times) [draw,circle,cross,minimum width=0.2 cm, below of=D] {}; 
\draw [->] (D) -- ([xshift=.0cm]times.north);
\draw [->] (times) -- ([xshift=.3cm]fcn.north);
\draw [->] (H+) -- ([xshift=.0cm]times.east);
\draw [->] (D) -| ([xshift=0.3cm]precoder.north);
\draw [->] (D) -| ([xshift=.2cm]sum_rate.north);
\end{tikzpicture}}}
    
    \subfigure[With WMMSE precoder]{\resizebox{.45\textwidth}{!}{\begin{tikzpicture}[cross/.style={path picture={ 
  \draw[black]
(path picture bounding box.south east) -- (path picture bounding box.north west) (path picture bounding box.south west) -- (path picture bounding box.north east);
}}]
\node (G) [rectangle] {$\mb{G}$};
\node (fcn) [box, below of=G, xshift=.3cm, yshift=-1.5cm] {FCN};
\draw [->] (G) -- ([xshift=-.3cm]fcn.north);
\node (phi) [rectangle, below of=fcn,yshift=-.7cm] {$\bs{\Phi}$};
\draw [->] (fcn) -- (phi);

\node (precoder) [box, right of=fcn, xshift=2cm, text width=2.5cm] {WMMSE Precoder (constant during one round of training)};
\draw [->] (G) -| ([xshift=-.3cm]precoder.north);
\draw [->] (phi) -| (precoder);
% \node (W) [rectangle, right of=precoder, xshift=1cm] {$\mb{V}$};
% \draw [->] (precoder) -- (W);

\node (sum_rate) [box, right of=precoder, xshift=2cm] {WSR};
\draw [->] (G) -| ([xshift=-.2cm]sum_rate.north);
\draw [->] (phi) -| (sum_rate);
% \draw [->] (precoder) -- (sum_rate);
\draw [->] (precoder) -- node [midway,above ] {$\mb{V}$} (sum_rate);

\node (D) [rectangle, above of=fcn, yshift=1.0cm, xshift=0.3cm] {$\mb{D}$};
\node (H+) [rectangle, above of=fcn, yshift=0.cm, xshift=1.2cm] {$\mb{H}^+$};
\node (times) [draw,circle,cross,minimum width=0.2 cm, below of=D] {}; 
\draw [->] (D) -- ([xshift=.0cm]times.north);
\draw [->] (times) -- ([xshift=.3cm]fcn.north);
\draw [->] (H+) -- ([xshift=.0cm]times.east);
\draw [->] (D) -| ([xshift=0.3cm]precoder.north);
\draw [->] (D) -| ([xshift=.2cm]sum_rate.north);
\end{tikzpicture}}}
    \end{footnotesize}
    \caption{Composition of the objective function. 
    With the MMSE precoder, the input of the FCN $\bs{\Gamma}$ comprises $\mb{G}$ and $\mb{D}\mb{H}^+$. 
    The channel $\mb{C}$ is determined by $\mb{G}$, $\mb{D}$ and $\bs{\Phi}$, see \eqref{eq:channel}.
    The MMSE precoding matrix $\mb{V}$ is computed with \eqref{eq:mmse}.
    The WSR is determined by $\mb{C}$ and $\mb{V}$ together.
    With the WMMSE precoder, the structure of the objective function is similar,
    but $\mb{V}$ is no more a function of $\mb{G}$, $\mb{D}$ and $\bs{\Phi}$,
    but a constant which is updated periodically.
    Therefore, we do not compute the derivative of $\mb{V}$.}
    \label{fig:objective function}
\end{figure}
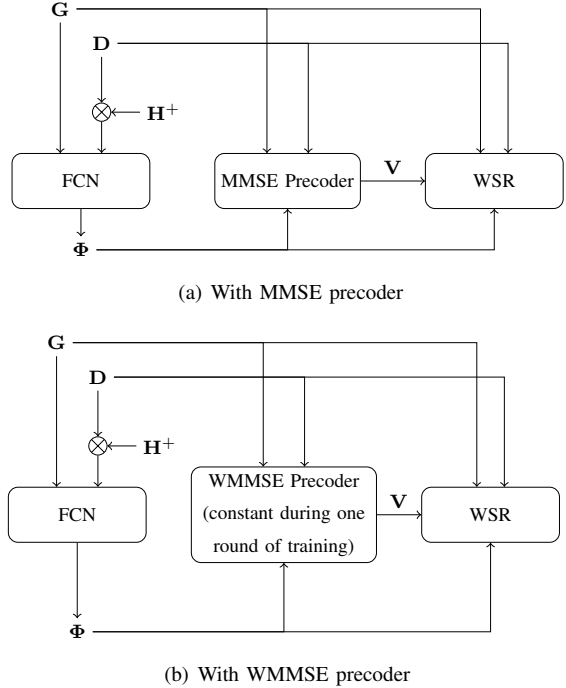

We can then compute $\frac{\partial f}{\partial \theta}$ to do gradient ascent to optimize $\p$ such that $f$ is maximized for both \eqref{eq:objective_mmse} and \eqref{eq:objective_wmmse}.
The compositions of the objective functions is illustrated in Fig.~\ref{fig:objective function}.
The Adam optimizer~\cite{kingma2014adam} is chosen to perform the optimization.

\rev{Since the gradient ascent improves the objective monotonically and the objective \gls{wsr} is bounded from above with a finite transmit power,
the iterative optimization must converge in the end of training.
In practice, we use stochastic gradient ascent instead of gradient ascent.
The gradient is computed for a randomly sampled batch of data rather than the whole data set.
After that, another batch with newly sampled data is used for the next gradient ascent step.
As a result,
we cannot guarantee monotonicity in the optimization.
However, the stochastic gradient ascent is a standard technique in machine learning and machine learning practice shows that if the data samples are identically distributed and the batch size is sufficiently large,
the stochastic gradient ascent shows solid convergence behavior,
which is also true in our case,
as will be shown in Section~\ref{sec:results}.}

\rev{Note that the proposed approach is an \emph{unsupervised} learning method.
In supervised learning, a target is provided for each input.
The optimizer tries to minimize the difference between the output of the neural network and the target (or label).
The loss function is usually defined as \gls{mse} (for continuous output) or cross entropy (for categorical output).
By minimizing the loss function, the neural network learns to reproduce the given target.
In our problem, the target is unknown and therefore not given.
By maximizing the objective function \eqref{eq:objective_mmse} or \eqref{eq:objective_wmmse},
the neural network learns an optimized mapping from channel features $\boldsymbol{\Gamma}$ to the \gls{ris} phase shift $\boldsymbol{\Psi}$ that maximizes the \gls{wsr}.
We believe that the unsupervised approach has a wider range of application since the target is unavailable for many difficult problems.
Besides, if the target is already available,
it is difficult to justify why we use machine learning to reproduce the target that is already known
(common reasons are, e.g., it is very time-consuming to obtain the target and the evaluation of the neural network is faster than the original algorithm~\cite{matthiesen2020globally}, or the purpose of machine learning is the predistortion of a distorted signal and the target (the original signal) is therefore known~\cite{wu2020residual}).
Nevertheless, it is of great practical importance and also very exciting if the neural network finds a good solution by itself without supervision.}

\section{Training Process Formulation and Implementation Considerations}
\label{sec:implementation}

\subsection{The Two-Phase Training}
\label{sec:two-phase}

As described in Section~\ref{sec:precoder}, the \gls{mmse} precoder is simple and differentiable and the \gls{wmmse} precoder has a higher performance at the cost of complexity and nondifferentiability. 
During the training, we apply the \gls{mmse} precoder and use \eqref{eq:objective_mmse} as the objective function in the first phase of training for a fast training 
and then switch to the \gls{wmmse} precoder and use \eqref{eq:objective_wmmse} as the objective function to further improve the \gls{wsr} in the second phase of training.

In the first phase, objective \eqref{eq:objective_mmse} is a function composition and its derivative can be computed with the chain rule.
The \gls{mmse} precoding matrix is computed with the current phase shifts $\bs{\Psi}$.
On the contrary, in the second phase, the precoding matrix $\mb{V}_{\text{WMMSE}}$ is nondifferentiable and is considered constant. 
Since $\mb{V}_{\text{WMMSE}}$ changes with the channel, when $\bs{\Psi}$ changes because of the training, $\mb{V}_{\text{WMMSE}}$ is no more valid.
Therefore, we update $\mb{V}_{\text{WMMSE}}$ every 10 epochs during training.

In the iterative \gls{wmmse} precoding algorithm (Algorithm~\ref{alg:wmmse}), the initial precoding matrix is arbitrary subject to the transmit power constraint.
The iteration is supposed to continue until convergence.
In our training, however, it is desirable that 
(a) we only run a few iterations such that the training time is reasonable,
(b) the updated precoding matrix is close to the precoding matrix from the previous update, 
such that a drastic change is avoided and the training process is more fluent.
With these considerations, we use the previous precoding matrix as the initial precoding matrix in the next update and run 5 iterations per update. 
In the first \gls{wmmse} precoding matrix update, the \gls{mmse} precoding matrix is applied.
Although the \gls{wmmse} precoding matrix is not updated in real time as the \gls{mmse} precoding matrix,
the change of the \gls{wmmse} is smooth and the training process is stable,
as will be shown in Section~\ref{sec:results}.

Besides the above-mentioned complexity consideration,
another important reason for applying \gls{mmse} precoder in the first phase of the training is that the \gls{wmmse} precoder selects users by weighting them with different $w_u$ in Algorithm~\ref{alg:wmmse}.
If the channel condition is poor and transmit power is insufficient following the water-filling principle~\cite{lozano2006optimum},
the \gls{wmmse} precoder selects good users and allocate no power to poor users,
which is the situation for many data samples in the beginning of training since the \gls{ris} is randomly initialized and cannot realize a good channel.
During the \gls{ris} optimization, since the poor user does not have any transmit power,
the \gls{ris} is optimized only for the good user and the poor user is ignored.
Therefore, the \gls{wmmse} precoder does not allocate any transmit power to the poor user in the next iteration because the channel to the good user is improved by the \gls{ris} but the channel to the poor user is not improved.
As a result, user selection under the initial poor channel quality may last until the end of training,
which results in a suboptimal performance.
On the contrary, the \gls{mmse} precoder treats the two users equally
and the \gls{ris} tries to improve the channel quality for both users.
In the end of the first phase,
channel quality is significantly improved compared to the initialization.
Therefore, user selection of the \gls{wmmse} precoder abandons less users,
which results in a high \gls{wsr} and better fairness.
The two-phase training is a reasonable choice especially when the initial channel quality is poor.

The proposed training is formulated in Algorithm~\ref{alg:training}.
\begin{algorithm}
\caption{FCN Training in two phases}
\label{alg:training}
\begin{algorithmic}
\Repeat
\State Compute the gradient of the objective \eqref{eq:objective_mmse} w.r.t. the neural network parameters
\State Perform a gradient ascent step with the Adam optimizer
\Until{WSR stops increasing}
\Repeat
\State Compute WMMSE precoding vectors $\mathbf{v}_u$ for all $u$ according to Algorithm~\ref{alg:wmmse}
\Repeat
\State Compute the gradient of the objective \eqref{eq:objective_wmmse} with current $\mathbf{v}_u$ w.r.t. the neural network parameters
\State Perform a gradient ascent step with the Adam optimizer
\Until{Preset number of iterations reached}
\Until{WSR stops increasing}
\end{algorithmic}
\end{algorithm}

\subsection{Solution for Discrete Phase Shifts}
\label{sec:discrete}

\rev{Although the above-described method optimizes the phase shifts of \gls{ris} antennas such that the \gls{wsr} is improved significantly,
a practical \gls{ris} is expected to have discrete phase shifts with a high granularity (e.g., $\pi$) rather than continuous phase shifts in order to reduce the hardware complexity and cost.
The phase shift of each antenna is chosen from a finite set of discrete phases $\boldsymbol{\Phi}_d$,
rather than a continuous value.
In this section, we propose a penalty term to convert continuous phase shifts to discrete phase shifts while keeping the performance loss as low as possible.
The penalty term is defined as the smallest Frobenius norm between continuous phase shift and the closest feasible discrete phase defined by $\boldsymbol{\Phi}_d$:
\begin{equation}
    p = \sqrt{\sum_n \left(\min_{\varphi_d \in \boldsymbol{\Phi}_d} \varphi_n - \varphi_d \right)}.
    \label{eq:penalty}
\end{equation}
The objective is defined as
\begin{equation}
    \sum_{u=1}^U \alpha_u \log_2\left( 1 + \frac{c_{uu}}{\sum_{v \neq u} c_{uv} + \frac{1}{\rho}} \right) - \pc p
    \label{eq:discrete_objective}
\end{equation}
where $\pc$ is the factor for the penalty term. The training process with discrete phase shifts is formulated as Algorithm~\ref{alg:discrete}.
\begin{algorithm}
\caption{FCN Training with Discrete Phase Shifts}
\label{alg:discrete}
\begin{algorithmic}
\Repeat
\State Compute the gradient of the objective \eqref{eq:objective_mmse} w.r.t. the neural network parameters
\State Perform a gradient ascent step with the Adam optimizer
\Until{WSR stops increasing}
\State Set $\pc = 0$
\Repeat
\Repeat
\State Compute WMMSE precoding vectors $\mathbf{v}_u$ for all $u$ according to Algorithm~\ref{alg:wmmse}
\Repeat
\State Compute the gradient of the objective \eqref{eq:discrete_objective} with current $\mathbf{v}_u$ w.r.t. the neural network parameters
\State Perform a gradient ascent step with the Adam optimizer
\Until{Preset number of iterations reached}
\Until{WSR stops increasing}
\State $\pc \leftarrow \pc + 0.05$
\Until{$p$ is smaller than a given threshold}
\end{algorithmic}
\end{algorithm}}

\subsection{Implementation}

We employ PyTorch~\cite{NEURIPS2019_9015} as the machine learning framework, 
which only supports real-valued inputs and parameters. 
Therefore, the channel features \eqref{eq:feature_definition} use amplitude and phase to describe a complex channel gain. 
On the contrary, since PyTorch 1.9, complex numbers are supported in the objective function, 
which significantly reduces the complexity in implementation of \eqref{eq:objective_mmse} and \eqref{eq:objective_wmmse}.
After implementation of the objective function, the gradient is computed with PyTorch's differentiation engine Autograd~\cite{NEURIPS2019_9015} without human intervention.
\section{Training and Evaluation Results}
\label{sec:results}

\begin{figure}[htbp]
    \centering
    \includegraphics[width=.45\textwidth, bb=0 0 549 456]{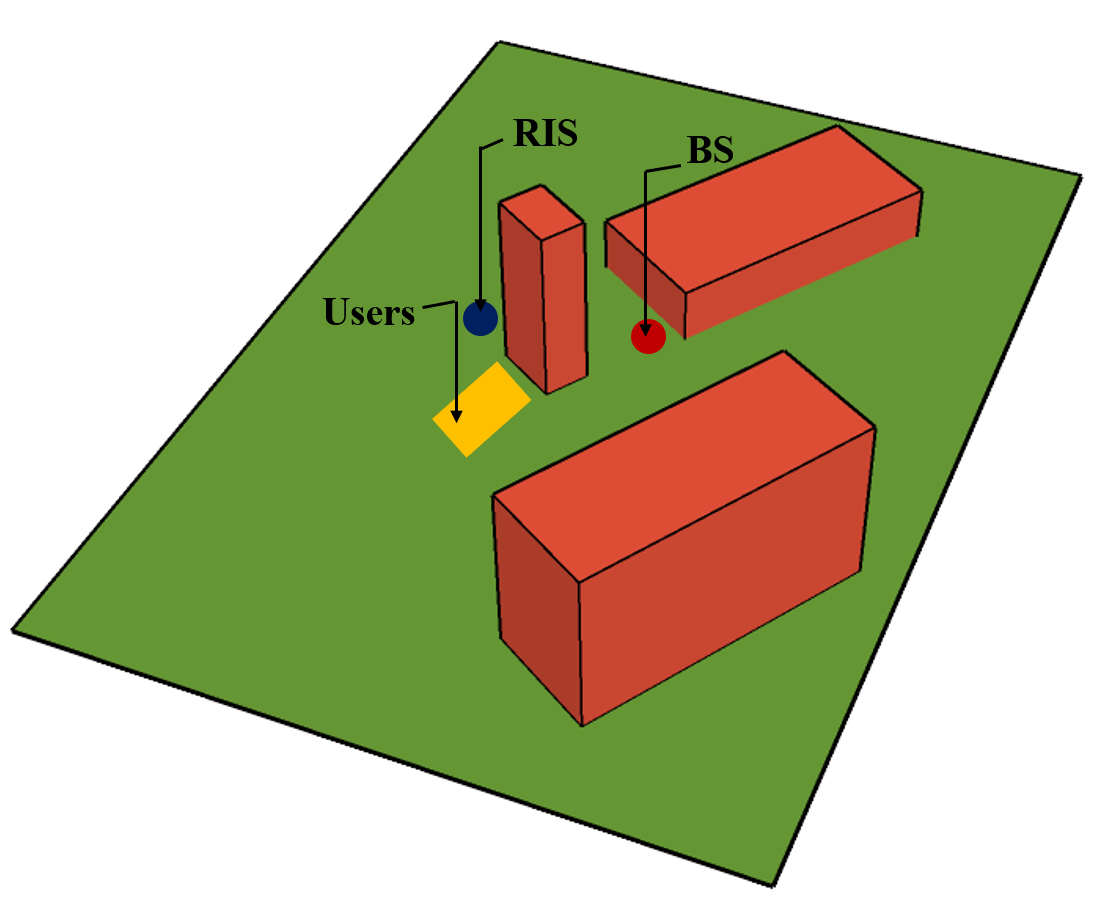}
    \caption{The considered scenario. 
    Positions of BS and RIS are given and positions of user pairs are generated randomly inside the yellow area.
    The LoS propagation path between BS and users is blocked by the high building in the middle.
    However, LoS propagation paths from BS to RIS and from RIS to users exist.
    A building behind the BS enables a strong reflection path from BS to RIS.
    A weak direct path is available through reflection on the building in the lower right corner.}
    \label{fig:scenario}
\end{figure}

In this section, we present the training and evaluation results.
We apply a ray tracing channel simulator in an urban environment~\cite{he2018design},
where mobile users are supposed to have high demand for data rates.
We generate channel from the \gls{bs} to the \gls{ris} $\mb{H}$,
channels from the \gls{ris} to the users $\mb{G}$ and the direct channels $\mb{D}$ of 5000 pairs of users,
whose positions are randomly generated with a minimum distance of 2 meters between them.
The scenario is shown in Fig.~\ref{fig:scenario},
where a \gls{los} propagation path between \gls{bs} and users is assumed to be unavailable due to the blockage by a building in the middle.
A weak direct path is available through reflection on the building in the lower right corner.
An \gls{ris} is equipped such that there are \gls{los} propagation paths from \gls{bs} to \gls{ris} and from \gls{ris} to users.
A building behind the \gls{bs} enables a strong reflection path from \gls{bs} to \gls{ris} such that channel matrix $\mb{H}$ has a rank higher\footnote{The rank of a matrix product is less than or equal to the rank of every matrix in the product. 
Since $\mb{G}$ is the channel matrix to two users at different positions, 
it has a rank of 2.
$\bs{\Phi}$ is a diagonal matrix and therefore has full rank.
Therefore, $\mb{H}$ has to have a rank higher than 1,
otherwise we have to use the weak direct channel $\mb{D}$,
which leads to a poor performance.} than 1.

Note that we assume the channel from \gls{bs} to \gls{ris} $\mb{H}$ is a constant,
which holds because both positions of \gls{bs} and \gls{ris} are fixed.
However, it requires per-\gls{bs}-\gls{ris}-pair channel measurement and training before deployment and when surrounding environment changes (e.g., a new building is built nearby),
which should be plausible since per-cell optimization is a common practice~\cite{shi2008mmse,baracca2012base,eckhardt2011vertical,zhang2021learning}.

The important parameters of scenario and model are presented in Table~\ref{tab:params}. 
If multiple parameter values are considered, the default value is marked with a star (*), 
which is always used without explicit specification.

\begin{table}[t]
    \centering
    \caption{Scenario and model parameters}
    \label{tab:params}
    \begin{tabularx}{.9\columnwidth}{XX}
        \hhline{==}
        Parameter & Value \\
        \hline
        Number of \gls{bs} antennas & 9 \\
        \gls{ris} size & (16, 16)*, (32, 16), (32, 32)\\
        Carrier frequency & 5.8 GHz\\
        Distance between adjacent antennas at \gls{bs} & 0.5 wave length\\
        Distance between adjacent antennas at \gls{ris} & 0.25 wave length\\
        Transmit \gls{snr} & $10^{11}$*, $1.1\times 10^{11}$, \dots $10^{12}$\\
        Weights of users & (0, 1), (0.25, 0.75), (0.5, 0.5)*, (0.75, 0.25), (1, 0)\\
        Filter size & (5, 5) for \gls{ris} size of (16, 16), (13, 13) otherwise\\
        Zero padding size & (2, 2) for \gls{ris} size of (16, 16), (6, 6) otherwise\\
        Number of convolutional layers & 8\\
        Dropout rate & 0.1 for \gls{ris} size of (16, 16), 0.35 otherwise\\
        Training rate with \gls{mmse} precoder & $10^{-4}$\\
        Epoches with \gls{mmse} precoder & 4000\\
        Training rate with \gls{wmmse} precoder & $2\times 10^{-6}$\\
        Updating interval with \gls{wmmse} precoder & 10 Epoches\\
        Epoches with \gls{wmmse} precoder & 4000\\
        Batch size & 256\\
        Optimizer & ADAM\\
        Number of data samples in training set & 5000\\
        Number of data samples in testing set & 1024\\
        \hhline{==}
    \end{tabularx}
\end{table}

\begin{figure}[htbp]
    \centering
    \resizebox{.45\textwidth}{!}{\input{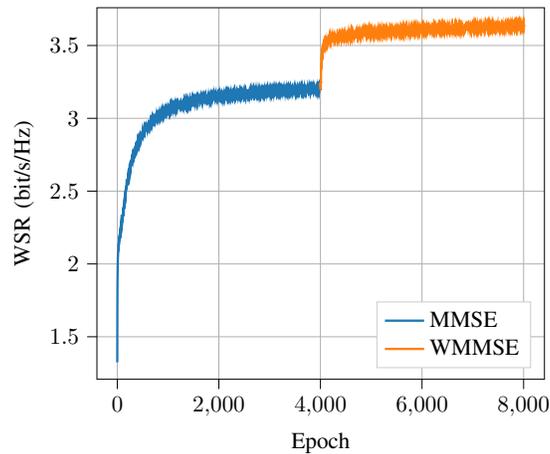}}
    \caption{Improvement of WSR in our proposed two-phase training protocol with pretraining with the MMSE precoder and ``booster'' training with the WMMSE precoder.}
    \label{fig:training}
\end{figure}

% Improvement of WSR in training
Fig.~\ref{fig:training} shows the improvement of the \gls{wsr} with two users, a \gls{tsnr} of $10^{11}$ and user weights of (0.5, 0.5) in training (therefore, the sum rate is the \gls{wsr} in the figure times 2). 
As explained in Section~\ref{sec:implementation},
we first use the \gls{mmse} precoder to train the model for 4000 epochs, 
then switch to the \gls{wmmse} precoder for further training.
It can be observed that the training has improved the \gls{wsr} significantly with the \gls{mmse} precoder and the \gls{wmmse} precoder outperforms the \gls{mmse} precoder.

\begin{figure}[htbp]
    \centering
    \resizebox{.45\textwidth}{!}{\input{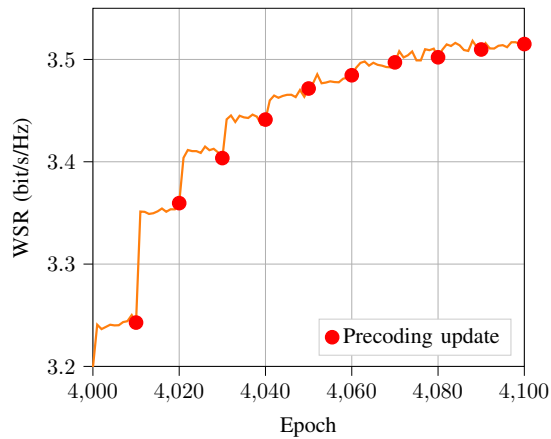}}
    \caption{Details to Fig. 6: Training with WMMSE precoder with update of precoding matrix every 10 epochs.}
    \label{fig:training_wmmse}
\end{figure}

Fig.~\ref{fig:training_wmmse} shows the first 100 epochs of the training with the \gls{wmmse} precoder (i.e., epochs 4000 - 4100 in Fig.~\ref{fig:training}).
Time points, where the precoding vectors are updated with the \gls{wmmse} precoder are marked with the red dots.
At the beginning of the training with the \gls{wmmse} precoder,
there is always a considerable performance improvement after the precoding vectors are updated.
The performance is then improved slightly due to the \gls{fcn} improvement.
The improvement slows down as the training carries on until convergence.

\begin{figure}[htbp]
    \centering
    \resizebox{.4\textwidth}{!}{\input{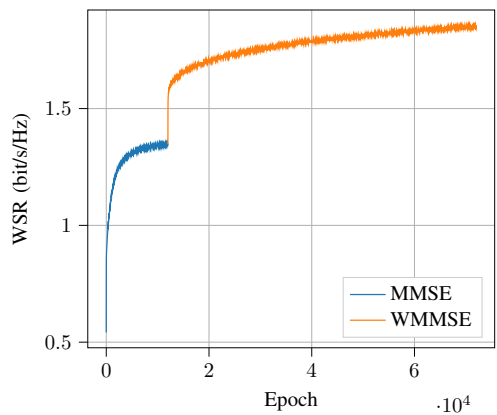}}
    \caption{Improvement of WSR in the two-phase training protocol for 4 users.}
    \label{fig:training_4users}
\end{figure}

\rev{Fig.~\ref{fig:training_4users} shows the \gls{wsr} improvement with four users and two-phase training.
Since the weights of all four users are 0.25,
the sum rate is the \gls{wsr} in the figure times 4.
Compare Fig.~\ref{fig:training} and Fig.~\ref{fig:training_4users},
we can observe that the scenario with four users achieves slightly better performance with longer training time.
The reason for the better performance is the higher potential to reuse the spectrum resource due to the larger number of users
whereas the reason for the longer training time is the more complicated objective function and neural architecture.
However, the advantage of four users compared to two users is only marginal.
As described in Section~\ref{sec:two-phase},
the \gls{wmmse} precoder allocates no transmit power to poor users when the available transmit power is insufficient following the water-filling principle.
In order to verify whether the marginal advantage of four users is due to the transmit power constraint,
we train a model with a \gls{tsnr} of $10^{13}$ while keeping other parameters unchanged\footnote{Note that this is not realistic assumption given a typical noise power.
The purpose of this assumption is only to validate whether the transmit power constraint prevents the four-user scenario from outperforming the two-user scenario significantly.}.}

\begin{figure}[htbp]
    \centering
    \resizebox{.4\textwidth}{!}{% This file was created with tikzplotlib v0.9.15.
\begin{tikzpicture}

\definecolor{color0}{rgb}{0.12156862745098,0.466666666666667,0.705882352941177}
\definecolor{color1}{rgb}{1,0.498039215686275,0.0549019607843137}
\definecolor{color2}{rgb}{0.172549019607843,0.627450980392157,0.172549019607843}

\begin{axis}[
legend cell align={left},
legend style={
  fill opacity=1,
  draw opacity=1,
  text opacity=1,
  at={(0.03,0.97)},
  anchor=north west,
  draw=white!80!black
},
tick align=outside,
tick pos=left,
x grid style={white!69.0196078431373!black},
xlabel={TSNR},
xmin=-0.38, xmax=1.38,
xtick style={color=black},
xtick={0,1},
xticklabels={$10^{11}$,$10^{13}$},
y grid style={white!69.0196078431373!black},
ylabel={Sum rate (bit/s/Hz)},
ymin=0, ymax=45,
ytick style={color=black}
]
\draw[draw=none,fill=color0] (axis cs:-0.3,0) rectangle (axis cs:-0.1,6.87);
\addlegendimage{ybar,ybar legend,draw=none,fill=color0}
\addlegendentry{Two users with FCN (ours)}

\draw[draw=none,fill=color0] (axis cs:0.7,0) rectangle (axis cs:0.9,19.21);
\draw[draw=none,fill=color1] (axis cs:-0.1,0) rectangle (axis cs:0.1,6.77);
\addlegendimage{ybar,ybar legend,draw=none,fill=color1}
\addlegendentry{Four users with BCD~\cite{guo2020weighted}}

\draw[draw=none,fill=color1] (axis cs:0.9,0) rectangle (axis cs:1.1,34.12);
\draw[draw=none,fill=color2] (axis cs:0.1,0) rectangle (axis cs:0.3,6.91);
\addlegendimage{ybar,ybar legend,draw=none,fill=color2}
\addlegendentry{Four users with FCN (ours)}

\draw[draw=none,fill=color2] (axis cs:1.1,0) rectangle (axis cs:1.3,38.54);
\draw (axis cs:-0.2,6.87) ++(0pt,3pt) node[
  anchor=south,
  text=black,
  rotate=0.0
]{6.87};
\draw (axis cs:0.8,19.21) ++(0pt,3pt) node[
  anchor=south,
  text=black,
  rotate=0.0
]{19.21};
\draw (axis cs:0,6.77) ++(0pt,3pt) node[
  anchor=south,
  text=black,
  rotate=0.0
]{6.77};
\draw (axis cs:1,34.12) ++(0pt,3pt) node[
  anchor=south,
  text=black,
  rotate=0.0
]{34.12};
\draw (axis cs:0.2,6.91) ++(0pt,3pt) node[
  anchor=south,
  text=black,
  rotate=0.0
]{6.91};
\draw (axis cs:1.2,38.54) ++(0pt,3pt) node[
  anchor=south,
  text=black,
  rotate=0.0
]{38.54};
\end{axis}

\end{tikzpicture}}
    \caption{Sum rate with 2 and 4 users and different algorithms.}
    \label{fig:4users_advantage}
\end{figure}
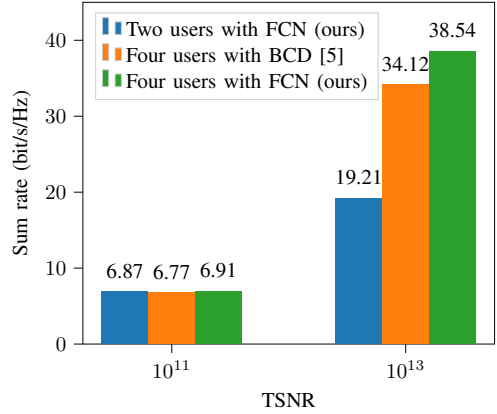

\rev{The comparison of achieved sum rate in the testing set (i.e., data that are not seen by the optimizer such that we can test the generality of the trained model) with equal user weights and different algorithms is shown in Fig.~\ref{fig:4users_advantage},
where we choose the \gls{bcd} algorithm in \cite{guo2020weighted} as a baseline.
It can be observed that all three setups achieve similar sum rates with the the \gls{tsnr} of $10^{11}$
but the two setups with four users significantly outperform the setup with two users with the \gls{tsnr} of $10^{13}$
and the proposed algorithm is better than the \gls{bcd} algorithm.
This result validates that the transmit power is the constraint that prevent the four-user scenario from outperforming the two-user scenario significantly.}

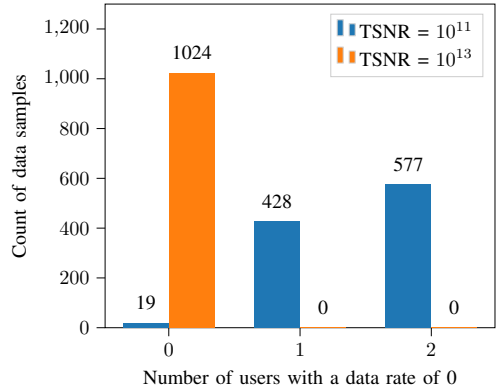
\begin{figure}[htbp]
    \centering
    \resizebox{.4\textwidth}{!}{% This file was created with tikzplotlib v0.9.15.
\begin{tikzpicture}

\definecolor{color0}{rgb}{0.12156862745098,0.466666666666667,0.705882352941177}
\definecolor{color1}{rgb}{1,0.498039215686275,0.0549019607843137}

\begin{axis}[
legend cell align={left},
  xtick={0, 1, 2},
legend style={
  fill opacity=1,
  draw opacity=1,
  text opacity=1,
  at={(0.97,0.97)},
  anchor=north east,
  draw=white!80!black
},
tick align=outside,
tick pos=left,
x grid style={white!69.0196078431373!black},
xmin=-0.485, xmax=2.485,
xtick style={color=black},
y grid style={white!69.0196078431373!black},
ylabel={Count of data samples},
xlabel={Number of users with a data rate of 0},
ymin=0, ymax=1300,
ytick style={color=black}
]
\draw[draw=none,fill=color0] (axis cs:-0.35,0) rectangle (axis cs:0,19);
\addlegendimage{ybar,ybar legend,draw=none,fill=color0}
\addlegendentry{TSNR = $10^{11}$}

\draw[draw=none,fill=color0] (axis cs:0.65,0) rectangle (axis cs:1,428);
\draw[draw=none,fill=color0] (axis cs:1.65,0) rectangle (axis cs:2,577);

\draw[draw=none,fill=color1] (axis cs:2.77555756156289e-17,0) rectangle (axis cs:0.35,1024);
\addlegendimage{ybar,ybar legend,draw=none,fill=color1}
\addlegendentry{TSNR = $10^{13}$}

\draw[draw=none,fill=color1] (axis cs:1,0) rectangle (axis cs:1.35,0);
\draw[draw=none,fill=color1] (axis cs:2,0) rectangle (axis cs:2.35,0);
\draw (axis cs:-0.175,19) ++(0pt,3pt) node[
  anchor=south,
  text=black,
  rotate=0.0
]{19};
\draw (axis cs:0.825,428) ++(0pt,3pt) node[
  anchor=south,
  text=black,
  rotate=0.0
]{428};
\draw (axis cs:1.825,577) ++(0pt,3pt) node[
  anchor=south,
  text=black,
  rotate=0.0
]{577};
\draw (axis cs:0.175,1024) ++(0pt,3pt) node[
  anchor=south,
  text=black,
  rotate=0.0
]{1024};
\draw (axis cs:1.175,0) ++(0pt,3pt) node[
  anchor=south,
  text=black,
  rotate=0.0
]{0};
\draw (axis cs:2.175,0) ++(0pt,3pt) node[
  anchor=south,
  text=black,
  rotate=0.0
]{0};
\end{axis}

\end{tikzpicture}}
    \caption{Number of users with a data rate of 0 in testing of the four-user scenario with the proposed algorithm.}
    \label{fig:num_without_datarate}
\end{figure}

\rev{Fig.~\ref{fig:num_without_datarate} provides further insights into the transmit power constraint,
where the numbers of data samples with different numbers of users with a data rate of 0 are shown for both transmit powers.
As shown in Table~\ref{tab:params}, 1024 data samples are tested in total.
We can observe that two users have a data rate of 0 in more than half (577) of all data samples with the \gls{tsnr} of $10^{11}$,
which makes the four-user scenario equivalent to the two-user scenario.
Therefore, the average performance difference between the two scenarios is very small.
On the contrary, no user has 0 data rate with the \gls{tsnr} of $10^{13}$.
This is because the resource is fully reused and the sum rate outperforms the two-user scenario as a result.}

% Rate region
To compare the proposed solution with other algorithms, 
we choose random \gls{ris} phase shifts and phase shifts optimized with the \gls{bcd} algorithm proposed as Algorithm~2 in \cite{guo2020weighted} as two baselines. 
\rev{The number of iterations is increased from 100 in the original paper to 5000 due to the larger number of \gls{ris} antennas.}
All three algorithms use the \gls{wmmse} precoder.
As before, the \gls{fcn} is evaluated with the testing data set.

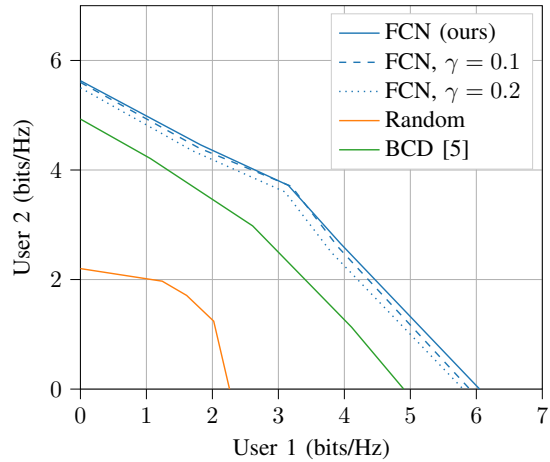
\begin{figure}[htbp]
    \centering
    \resizebox{.45\textwidth}{!}{% This file was created with tikzplotlib v0.9.15.
\begin{tikzpicture}

\definecolor{color0}{rgb}{0.12156862745098,0.466666666666667,0.705882352941177}
\definecolor{color1}{rgb}{1,0.498039215686275,0.0549019607843137}
\definecolor{color2}{rgb}{0.172549019607843,0.627450980392157,0.172549019607843}

\begin{axis}[
legend cell align={left},
legend style={fill opacity=1, draw opacity=1, text opacity=1, draw=white!80!black},
tick align=outside,
tick pos=left,
x grid style={white!69.0196078431373!black},
xlabel={User 1 (bits/Hz)},
xmajorgrids,
xmin=-0, xmax=7,
xtick style={color=black},
y grid style={white!69.0196078431373!black},
ylabel={User 2 (bits/Hz)},
ymajorgrids,
ymin=-0, ymax=7,
ytick style={color=black}
]
\addplot [semithick, color0]
table {%
0 5.63
1.82 4.46
3.15 3.72
3.94 2.66
6.05 0
};
\addlegendentry{FCN (ours)}
\addplot [semithick, color0, dashed]
table {%
0 5.6
1.8 4.4
3.2 3.7
3.9 2.6
5.9 0
};
\addlegendentry{FCN, $\gamma=0.1$}
\addplot [semithick, color0, dotted]
table {%
0 5.5
1.7 4.35
3.1 3.6
3.8 2.5
5.8 0
};
\addlegendentry{FCN, $\gamma=0.2$}
\addplot [semithick, color1]
table {%
0 2.2
1.24 1.97
1.61 1.71
2.02 1.24
2.26 0
};
\addlegendentry{Random}
\addplot [semithick, color2]
table {%
0 4.93
1.06 4.21
2.61 2.98
4.11 1.13
4.9 0
};
\addlegendentry{BCD \cite{guo2020weighted}}
\end{axis}

\end{tikzpicture}}
    \caption{Average rate regions of baselines and proposed algorithm.}
    \label{fig:rate_regions}
\end{figure}

Fig.~\ref{fig:rate_regions} shows the rate regions of the three algorithms.
Each curve is obtained with the user weights of (0, 1), (0.25, 0.75), (0.5, 0.5), (0.75, 0.25) and (1, 0) and
every point is the average of performances of 1024 channel samples.
We can see that both the \gls{bcd} algorithm and the \gls{fcn} improve the data rates of two users significantly from the random initialization.
The \gls{fcn} outperforms the \gls{bcd} algorithm clearly.
Besides, although training \gls{fcn} and running \gls{bcd} algorithm have time consumption in the same order of magnitude,
training the \gls{fcn} is only done once before the application.
Testing \gls{fcn} with 1024 channel samples takes merely a few seconds 
while running the \gls{bcd} algorithm takes more than 4 hours with a 12-core CPU 
(channel samples are processed in parallel with the \emph{parfor} loop in MATLAB).
This makes the proposed approach much more promising to be deployed in the near future
for real-time applications given hardware with comparable performance as of today.
\rev{Besides, Fig.~\ref{fig:rate_regions} also shows the achieved sum rate with random channel estimate error,
which is assumed to be an \gls{iid} Gaussian random error.
The ratio between standard deviation of the Gaussian random error and the mean channel gain is denoted as $\gamma$.
We can observe that the \gls{fcn} is robust against a channel estimate error
partially because of the dropout layers,
which add noise to inputs of each convolutional layer in the training and therefore enhance robustness of the neural work.
This result shows that the proposed solution does not require very precise channel estimation in order to work,
which would be very challenging in \gls{ris}-aided systems.}

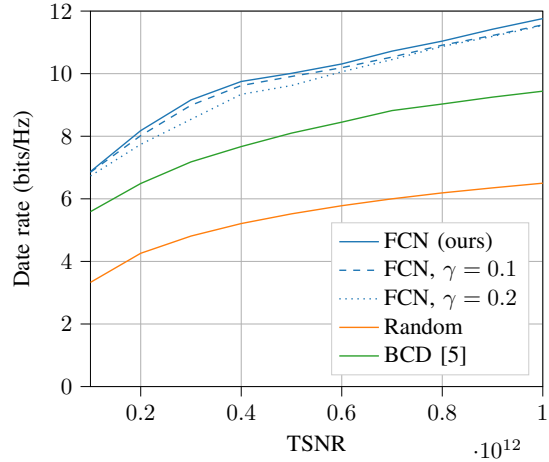
\begin{figure}[htbp]
    \centering
    \resizebox{.45\textwidth}{!}{% This file was created with tikzplotlib v0.9.15.
\begin{tikzpicture}

\definecolor{color0}{rgb}{0.12156862745098,0.466666666666667,0.705882352941177}
\definecolor{color1}{rgb}{1,0.498039215686275,0.0549019607843137}
\definecolor{color2}{rgb}{0.172549019607843,0.627450980392157,0.172549019607843}

\begin{axis}[
legend cell align={left},
legend style={
  fill opacity=1,
  draw opacity=1,
  text opacity=1,
  at={(0.97,0.03)},
  anchor=south east,
  draw=white!80!black
},
tick align=outside,
tick pos=left,
x grid style={white!69.0196078431373!black},
xlabel={TSNR},
xmajorgrids,
xmin=100000000000, xmax=1000000000000,
xtick style={color=black},
y grid style={white!69.0196078431373!black},
ylabel={Date rate (bits/Hz)},
ymajorgrids,
ymin=0, ymax=12,
ytick style={color=black}
]
\addplot [semithick, color0]
table {%
100000000000 6.87
200000000000 8.18
300000000000 9.16
400000000000 9.75
500000000000 10.01
600000000000 10.31
700000000000 10.72
800000000000 11.04
900000000000 11.42
1000000000000 11.76
};
\addlegendentry{FCN (ours)}
\addplot [semithick, color0, dashed]
table {%
100000000000 6.85
200000000000 8.02
300000000000 8.99
400000000000 9.62
500000000000 9.91
600000000000 10.19
700000000000 10.53
800000000000 10.91
900000000000 11.22
1000000000000 11.56
};
\addlegendentry{FCN, $\gamma=0.1$}
\addplot [semithick, color0, dotted]
table {%
100000000000 6.73
200000000000 7.74
300000000000 8.54
400000000000 9.34
500000000000 9.62
600000000000 10.06
700000000000 10.45
800000000000 10.87
900000000000 11.19
1000000000000 11.54
};
\addlegendentry{FCN, $\gamma=0.2$}
\addplot [semithick, color1]
table {%
100000000000 3.33
200000000000 4.26
300000000000 4.81
400000000000 5.21
500000000000 5.52
600000000000 5.78
700000000000 6
800000000000 6.19
900000000000 6.35
1000000000000 6.5
};
\addlegendentry{Random}
\addplot [semithick, color2]
table {%
100000000000 5.59
200000000000 6.49
300000000000 7.18
400000000000 7.67
500000000000 8.1
600000000000 8.45
700000000000 8.82
800000000000 9.03
900000000000 9.25
1000000000000 9.44
};
\addlegendentry{BCD~\cite{guo2020weighted}}
\end{axis}

\end{tikzpicture}}
    \caption{Sum rates of baselines and proposed algorithm with different TSNRs.}
    \label{fig:tsnr}
\end{figure}

% WSR with different TSNR
Fig.~\ref{fig:tsnr} shows the sum rate with the same user weights (0.5, 0.5) and different \gls{tsnr}.
We can see that higher \gls{tsnr} (i.e., transmit power) increases the data rate considerably
and the \gls{ris} configured by the \gls{fcn} outperforms random phase shifts and phase shifts optimized with the \gls{bcd} algorithm for all tested \glspl{tsnr}.
It is also to note that the performance of the \gls{fcn} is robust against random channel estimation errors.

In the evaluation, we have obtained a \gls{wsr} of 6.87~bit/s/Hz for \gls{tsnr}=$10^{11}$.
Compared to the \gls{wsr} at the end of training (7.26~bit/s/Hz), 
the \gls{wsr} in the evaluation is roughly 5\% less than the \gls{wsr} in the training,
which suggests a low overfitting level and a good generalization of the trained model.

% Furthermore, comparing Fig.~\ref{fig:training} and Fig.~\ref{fig:tsnr},
% we can see that the final sum rate in training (i.e., two times (because each user has a weight of 0.5)
% the y coordinate of the last point of the curve in Fig.~\ref{fig:training}, which is 7.26 bit/s/Hz)
% and the sum rate in testing (i.e., the y coordinate of the green curve when \gls{tsnr}=$10^{11}$, which is 6.87 bit/s/Hz) are similar (the sum rate in training is roughly 5\% higher than the sum rate in testing),
% which suggests a low overfitting level and a good generalization of the trained model.

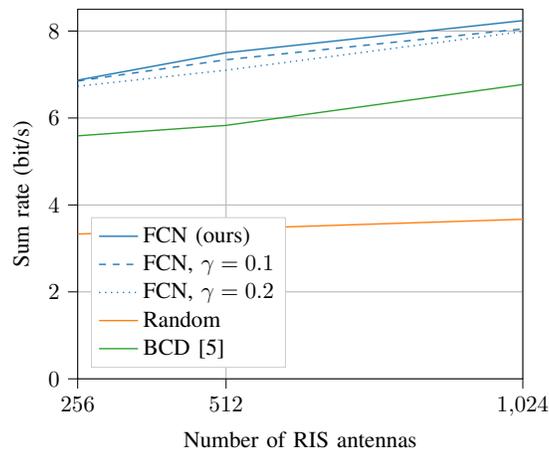
\begin{figure}[htbp]
    \centering
    \resizebox{.45\textwidth}{!}{% This file was created with tikzplotlib v0.9.15.
\begin{tikzpicture}

\definecolor{color0}{rgb}{0.12156862745098,0.466666666666667,0.705882352941177}
\definecolor{color1}{rgb}{1,0.498039215686275,0.0549019607843137}
\definecolor{color2}{rgb}{0.172549019607843,0.627450980392157,0.172549019607843}

\begin{axis}[
legend cell align={left},
legend style={
  fill opacity=1,
  draw opacity=1,
  text opacity=1,
  at={(0.03,0.03)},
  anchor=south west,
  draw=white!80!black
},
tick align=outside,
tick pos=left,
xtick={256,512,1024},
x grid style={white!69.0196078431373!black},
xlabel={Number of RIS antennas},
xmajorgrids,
xmin=256, xmax=1024,
xtick style={color=black},
y grid style={white!69.0196078431373!black},
ylabel={Sum rate (bit/s)},
ymajorgrids,
ymin=0, ymax=8.5,
ytick style={color=black}
]
\addplot [semithick, color0]
table {%
256 6.87
512 7.5
1024 8.24
};
\addlegendentry{FCN (ours)}
\addplot [semithick, color0, dashed]
table {%
256 6.85
512 7.34
1024 8.05
};
\addlegendentry{FCN, $\gamma=0.1$}
\addplot [semithick, color0, dotted]
table {%
256 6.73
512 7.1
1024 7.99
};
\addlegendentry{FCN, $\gamma=0.2$}
\addplot [semithick, color1]
table {%
256 3.33
512 3.45
1024 3.67
};
\addlegendentry{Random}
\addplot [semithick, color2]
table {%
256 5.59
512 5.83
1024 6.77
};
\addlegendentry{BCD~\cite{guo2020weighted}}
\end{axis}

\end{tikzpicture}}
    \caption{Sum rates of baselines and proposed algorithm with different numbers of RIS antennas.}
    \label{fig:num_antennas}
\end{figure}

% WSR with different numbers of antennas
Fig.~\ref{fig:num_antennas} shows the sum rates of different approaches with different number of \gls{ris} antennas.
It is to see that our proposed solution outperforms the two baselines with all tested numbers of antennas
and is robust against random channel estimation errors.
Besides, although the \gls{fcn} requires more training time due to bigger filter sizes (see TABLE~\ref{tab:params}) and more \gls{ris} antennas,
the evaluation of the \gls{fcn} is within two minutes even for 1024 antennas on the author's two-year-old laptop with Intel's 8th generation i7 CPU (evaluation is run on the CPU) and 8 GB RAM,
as shown in TABLE~\ref{tab:time_consumption}.
In contrast, the \gls{bcd} algorithm takes more than 3 days to optimize the \gls{ris} configuration with 1024 antennas on the server with a 12-core CPU,
which makes it almost impractical for real-time deployment.

\begin{table}[htbp]
    \centering
    \caption{Time consumption of different approaches}
    \label{tab:time_consumption}
    \begin{tabular}{llll}
    \hhline{====}
        \# of antennas & 256 & 512 & 1024 \\
    \hline
        Random & Seconds & Seconds & Seconds\\
        BCD & 5 hours & 1 day & 3 days \\
        FCN training & 5 hours & 2 days & 3 days \\
        FCN evaluation & Seconds & Seconds & Minutes \\
    \hhline{====}
    \end{tabular}
\end{table}

\begin{figure}[htbp]
    \centering
    % \subfigure[TSNR=$10^{11}$]{
    \resizebox{.45\textwidth}{!}{\input{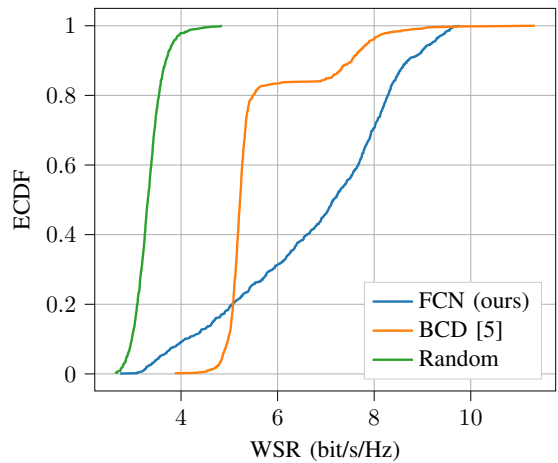}}
    % }
    % \subfigure[TSNR=$10^{12}$]{
    % \resizebox{.45\textwidth}{!}{\input{figs/ecdf_1e12}}
    % }
    \caption{ECDF of baselines and proposed algorithm.}
    \label{fig:ecdf}
\end{figure}

% ECDF
Fig.~\ref{fig:ecdf} shows the \gls{ecdf} of the sum-rate with 256 \gls{ris} antennas, a \gls{tsnr} of $10^{11}$ and the user weights of (0.5, 0.5),
where we can confirm the advantage of the proposed algorithm again.
It is also to note that the \gls{ecdf} of the proposed algorithm is flatter than the other two algorithms
and the outage probability that the \gls{wsr} is less than a given threshold is higher with the proposed algorithm than with the \gls{bcd} algorithm proposed in \cite{guo2020weighted},
suggesting that the performance depends more heavily on the channels.
The proposed algorithm only brings significantly advantage with favorable channels.
Therefore, user selection is crucially important for the best performance,
which remains an open problem.

\begin{figure}[htbp]
    \centering
    \resizebox{.45\textwidth}{!}{\input{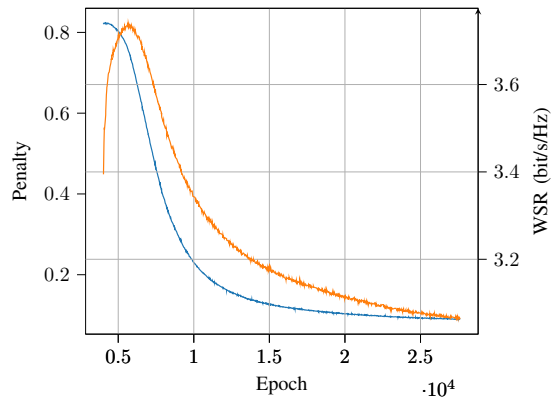}}
    \caption{Development of penalty (blue) and WSR (orange) during training during the training phase with the WMMSE precoder. A lower penalty suggests that the output phase shifts are closer to the predefined discrete phase shifts. The WSR reduces as the FCN output is being discretized. The purpose of the training is to minimize the WSR loss during the discretization.}
    \label{fig:penalty}
\end{figure}

\rev{Finally, we evaluate the performance with discrete phase shifts with a granularity of $\pi$. 
Fig.~\ref{fig:penalty} shows the penalty (defined in \eqref{eq:penalty}) development during training with the \gls{wmmse} precoder.
Algorithm~\ref{alg:discrete} effectively reduces the difference between the output phase shift and the predefined discrete phase shifts.
The \gls{wsr} reduces as the \gls{fcn} output is being discretized.
After training, we force the output phase to be one of the predefined phases and compare the result between the model without discretization (i.e., with Algorithm~\ref{alg:training}) and the model with discretization (i.e., with Algorithm~\ref{alg:discrete}),
which realize a sum rate of 5.72~bit/s/Hz and 6.05~bit/s/Hz, respectively.
This result suggests that the proposed discretization method in Section~\ref{sec:discrete} realizes only a marginal improvement compared to brutal discretization.
To the authors' best knowledge,
it remains an open problem to obtain the global optimum of a high dimensional combinatorial optimization problem.}

\rev{Besides the simpler \gls{ris} hardware,
another benefit of the discrete phase shift is the small amount of controlling signal.
With the default assumption of number of \gls{ris} antennas
and two discrete phase shifts,
which can be represented by one bit,
the \gls{ris} controlling signal has a data amount of 256~bit,
which can be easily transmitted within the next-generation wireless communication systems.
}

\section{Conclusions}
\label{sec:conclusion}

We proposed an \gls{fcn} based solution for spatial multiplexing enabled by a \gls{ris} in this paper.
The \gls{fcn} is a widely applied neural network architecture for computer vision but has not yet been used for \gls{ris} configuration.
The rectangular shape of the \gls{ris} and the spatial correlation of channels on adjacent antennas of the \gls{ris} encourage us to apply the \gls{fcn} to the \gls{ris} configuration.
We use the simple and differentiable \gls{mmse} precoder for pretraining 
and the iterative and complex \gls{wmmse} precoder with periodically updated precoding vectors to fine-tune the model.
A set of channel features is designed to include both cascaded channels via the \gls{ris} and the direct channel.
We use a ray-tracing simulator to generate channel models with spatial correlation.
Evaluation results show that the proposed solution achieves higher performance and also shows faster evaluation speed than the baselines. 
Therefore, it can be better scaled to a large number of antennas.
The dropout layers between the convolutional layers of the \gls{fcn} reduce overfitting.
A penalty term is introduced to discretize \gls{fcn} output such \rev{that the proposed method can be applied to \glspl{ris} with discrete phase shifts.}

It is shown that the \gls{wsr} depends strongly on the wireless channels.
As future works, it is desirable to find a criterion to optimally decide which users are selected to share the same resource block, 
which can significantly improve the average performance.
Furthermore, more advanced techniques of interference management, 
such as nonorthogonal multiple access (NOMA) and rate splitting, 
can also be combined with \gls{ris} and may achieve a better performance than treating interference as noise.

The source code and channel data of this paper will be provided for public access if the article is accepted.

\section*{Acknowledgement}

The results with random \gls{ris} phase shifts and with \gls{bcd} algorithm are obtained with the open-source code generously shared by authors of~\cite{guo2020weighted} under \url{https://github.com/guohuayan/WSR_maximization_for_RIS_system}.

The authors would like to thank Mrs. X. Shan for providing the ray-tracing channel models,
Dr. K. L. Besser for the fruitful discussion on the combinarotial optimization
and Mr. R. Wang for assisting training and simulation.

% \appendices

% \section{}

\ifCLASSOPTIONcaptionsoff
  \newpage
\fi


\printbibliography

\end{document}